\documentclass[12pt]{amsart}
\usepackage{amsmath,amssymb,esint,enumerate,graphicx,hyperref}
\usepackage[sort&compress,numbers]{natbib}

\hypersetup{pdftex,colorlinks=true,allcolors=blue}
\usepackage[all]{hypcap}

\usepackage[foot]{amsaddr}
\usepackage[sc]{mathpazo}

\newcommand\ack{\subsection*{Acknowledgment}}

\DeclareMathAlphabet\mathsfbi{T1}{phv}{b}{it}

\numberwithin{equation}{section}

\newcommand\BV{\boldsymbol} % Vector bold
\newcommand\BM{\mathsfbi} % Matrix and tensor bold

\newcommand\dif{\:\!\mathrm{d}}
\newcommand\deriv[2]{\frac{\mathrm{d} #1}{\mathrm{d} #2}}
\newcommand\parderiv[2]{\frac{\partial #1}{\partial #2}}

\newcommand\coll{\mathcal C}

\newcommand\EE{\mathbb E}
\newcommand\RR{\mathbb R}

\newcommand\cS{\mathcal S}
\newcommand\cA{\mathcal A}

\newcommand\cI{\mathcal I}

\newcommand\collmap{C}

\def\HS{{H\!S}}

\begin{document}

\author[Rafail V. Abramov]{Rafail V. Abramov}

\address{Department of Mathematics, Statistics and Computer Science,
University of Illinois at Chicago, 851 S. Morgan st., Chicago, IL 60607}

\email{abramov@uic.edu}

\title{Creation of turbulence in polyatomic gas flow via
an~intermolecular potential}

\begin{abstract}
We develop a tractable interaction model for a polyatomic gas, whose
kinetic equation combines a Vlasov-type mean field forcing due to an
intermolecular potential, and a Boltzmann-type collision integral due
to rotational interactions. We construct a velocity moment hierarchy
for the new kinetic equation, and find that, under the high Reynolds
number condition, the pressure equation becomes decoupled from the
angular momentum and stress. For the heat flux, we propose a novel
closure by prescribing the specific heat capacity of the gas flow.
Setting the specific heat capacity to that of a constant-pressure
process leads to the system of equations for a balanced flow, where
the momentum transport equation contains the mean field forcing, which
is an averaged effect of the intermolecular potential. Remarkably, the
balanced flow equations do not contain any information about internal
thermodynamic properties of the gas, and are thereby applicable to a
broad range of different gases. We conduct numerical simulations for
an air-like gas at normal conditions in the inertial flow regime,
where the pressure is constant throughout the domain. We find that the
presence of the intermolecular potential produces a distinctly
turbulent flow, whose time-averaged Fourier spectra of the kinetic
energy and temperature exhibit Kolmogorov's power decay.
\end{abstract}

\maketitle

\section{Introduction}

While turbulent motions in a liquid have first been documented by
Leonardo da Vinci, it appears that \citet{Bou} presented the first
systematic account of turbulence in the scientific literature. Six
years later, \citet{Rey83} demonstrated that an initially laminar flow
of a liquid consistently develops turbulent motions whenever the high
Reynolds number condition is satisfied. Almost sixty years later,
Kolmogorov \citep{Kol41a,Kol41b,Kol41c} and \citet{Obu41} observed that the
time-averaged Fourier spectra of the kinetic energy of an atmospheric
wind possess a universal decay structure, corresponding to the inverse
five-thirds power of the Fourier wavenumber. Numerous attempts to
explain the physical nature of turbulence have been made throughout
the twentieth century
\citep{Ric26,Tay,Tay38,KarHow38,Pra38,Obu49,Cha49,Cor51,Kol62,Obu62,Kra66a,Kra66b,Saf67,Saf70,Man74},
yet none successfully. Until recently, the physics of turbulence
formation in an initially laminar flow, as well as the origin of the
power scaling of turbulent kinetic energy spectra, remained unknown.
As an example, one can refer to relatively recent works
\citep{AviMoxLozAviBarHof,KhaAnwHasSan}, where turbulent-like motions
in a numerically simulated flow had to be created artificially by
deliberate perturbations. In reality, turbulence emerges spontaneously
by itself (even if all reasonable measures were taken to preserve the
laminarity of the flow, e.g. Reynolds' experiment), which was the
primary reason why this peculiar phenomenon attracted world-wide
scientific interest to begin with.

In our recent works \citep{Abr22,Abr23}, we considered a system of
many particles, interacting solely via a repelling short-range
potential $\phi(r)$. In the limit of infinitely many such particles,
we obtained, via the standard Bogoliubov--Born--Green--Kirkwood--Yvon
(BBGKY) formalism~\citep{Bog,BorGre,Kir}, a Vlasov-type
equation~\citep{Vla} for the mass-weighted distribution density of a
single particle. This equation contained the mean field potential
$\bar\phi$, which was the result of the average effect of the
short-range potential $\phi$. Then, we computed the usual hierarchy of
the transport equations for the velocity moments, closed it under the
infinite Reynolds number assumption and arrived at a system of
equations for a balanced compressible gas flow.

In \citep{Abr22}, we numerically simulated these equations in the
inertial flow regime through a straight pipe, with the interaction
potential of hard spheres with the mass and diameter corresponding to
those of argon. For the initial condition of that simulation, we
selected a straight laminar Bernoulli jet, which happens to be a
steady state in the conventional Euler equations. We observed,
however, that in our simulation such a jet quickly developed into a
fully turbulent flow. We also examined the Fourier spectrum of the
kinetic energy of the simulated flow, and found that its time average
decayed with the rate of inverse five-third power of the wavenumber,
which corresponded to the famous Kolmogorov spectrum.  In
\citep{Abr23}, we extended the framework of \citep{Abr22} onto a gas
flow under strong gravity acceleration, which, on a large scale, is
effectively two-dimensional. We simulated the two-dimensional
equations in the inertial and cyclostrophic flow regimes, and found
that turbulent motions also appear in an initially laminar flow.

In the past, a similar approach, originating from the basic principles
of kinetic theory, was undertaken by \citet{Tsu}, who attempted to
explain the creation of turbulence via long-range correlations between
molecules. However, Tsug\'e's result was restricted to incompressible
flow, while, from what we found thus far, it appears that density
fluctuations are instrumental in the creation of turbulent dynamics.
We have considered long-range interaction effects in our past work
\citep{Abr20}, however, from what we later found in
\citep{Abr22,Abr23}, it appears that even a short-range hard sphere
potential is capable of creating turbulence via the mean field effect.

The goal of the current work is to extend the turbulent framework of
\citep{Abr22} and \citep{Abr23} onto polyatomic gases. The main
challenge with a polyatomic gas is the corresponding kinetic model of
interactions between its molecules. Unlike monatomic gases (which can,
in many practical situations, be approximated via hard spheres),
molecules of polyatomic gases possess rotational degrees of freedom,
which store the momentum and energy just like the translational
degrees of freedom do. When two such molecules interact, the momentum
and energy are exchanged, generally, between linear and angular
velocities of both molecules in a complex manner. In order to derive
transport equations for a polyatomic gas from kinetic theory, these
complex interactions must be described, to a necessary extent, in the
context of the kinetic model of intermolecular collisions.

The work is organized as follows. In Section~\ref{sec:collisions} we
propose a tractable kinetic model of interactions of polyatomic gas
molecules, which combines a deterministic potential with random
collisions. In Section~\ref{sec:dyn_sys} we introduce a dynamical
system of many molecules which interact according to the
aforementioned model, and construct its corresponding forward
Kolmogorov equation. In Section~\ref{sec:boltzmann-vlasov} we compute
the equation for the statistical distribution of a single molecule,
which combines a potential forcing with a Boltzmann-like collision
integral, and derive the corresponding system of transport equations
of the velocity moments. In Section~\ref{sec:closure} we decouple the
pressure equation from the angular momentum and stress via the high
Reynolds number condition, and implement a closure for the divergence
of the heat flux by prescribing the specific heat capacity of the
process. For the specific heat capacity corresponding to that of a
process at constant pressure (that is, Charles' law of classical
thermodynamics), we arrive at the same system of balanced flow
equations as in our recent works \citep{Abr22,Abr23} for monatomic
gases. In Section~\ref{sec:numerical} we show the results of numerical
simulations of a turbulent air flow through a straight pipe. We
demonstrate that, in the presence of an intermolecular potential, the
flow becomes turbulent, and the time averages of the Fourier spectra
of its kinetic energy and temperature exhibit a power decay.
Section~\ref{sec:summary} summarizes the results of our work.

\section{The kinetic model of interactions of polyatomic molecules}
\label{sec:collisions}

At the first step, we need to formulate a tractable mathematical model
of a polyatomic gas on the molecular level.
Before proceeding further, however, we have to emphasize that the
actual interaction of atoms is a rather complicated process, which
involves quantum-mechanical effects such as the Pauli exclusion.
Therefore, what follows has to be interpreted strictly in the context
of a vastly simplified, i.e. ``mechanical'', model of atomic and
molecular interaction, whose purpose is to provide an appropriate
formalism which connects the kinetic formulation of the problem and
its fluid-mechanical approximation, consistent with fundamental
properties of interaction such as the conservation of energy.

In addition to the three usual translational degrees of freedom, each
molecule of a polyatomic gas also possesses multiple rotational
degrees of freedom, which are capable of storing the momentum and
energy. When such molecules interact, they exchange the momentum and
energy between all degrees of freedom of both molecules. Unlike
monatomic gases, the main challenge with a polyatomic gas is to
describe its molecular interactions, which can roughly be separated
into two issues.

\subsection{Excessive complexity of a detailed molecular interaction}

The first issue to consider with a polyatomic gas, is that the exact
description of the collision mechanics of two objects beyond that of a
pair of hard spheres is excessively complicated for a practical
treatment in the context of a kinetic equation. For example, even if
two hard spheres are replaced with two rigid ellipsoids (which is a
simple topological deformation of a sphere), the collision mechanics
already become too complicated to use in a Boltzmann-like collision
integral. In particular, not only the actual mechanics of
transformation of linear and angular velocities are complex, but even
the collision detection criterion is highly nontrivial (as an example,
refer to \citet{JiaChoMouWan} and references therein).

\subsection{The effect of Heisenberg's uncertainty principle}

Even if one develops a tractable model of mechanical collisions of two
molecules beyond that of hard spheres, there is another difficulty
which renders the utility of such model somewhat questionable. Due to
Heisenberg's uncertainty principle, gas molecules cannot be regarded
as fully classical objects, especially if the rotational motion of a
molecule is involved. Let us look at the following crude estimates of
uncertainties associated with the translational and rotational motion:
\begin{itemize}
\item A molecule of nitrogen (which we take as an example because it
  is the primary component of air) has the mass of $4.65\cdot
  10^{-26}$ kg, which means that the product of the uncertainties in
  the velocities and coordinates is about $10^{-9}$ m$^2$/s (the
  Planck constant divided by the mass). The typical mean free path
  between collisions at normal conditions is around $7\cdot 10^{-8}$
  m, while the average speed of a molecule at normal temperature is
  around $500$ m/s. Their product is $\sim 10^{-5}$, which gives the
  relative uncertainty, associated with translational motion, of $\sim
  10^{-4}$, or 0.01\%.
\item At the same time, the size of a molecule of nitrogen is $4\cdot
  10^{-10}$ m, which means, that, during the rotation, the two atoms
  move about each other on this scale. Now, since the kinetic energy
  of motion is distributed uniformly across translational and
  rotational degrees of freedom, we have to assume that the atoms of
  nitrogen move about each other within the molecule at roughly same
  speeds, that is, $\sim 500$ m/s. Their product is $\sim 10^{-7}$,
  which gives the relative uncertainty, associated with rotational
  motion, of $\sim 10^{-2}$, or 1\%.
\end{itemize}
As we can see, the relative uncertainty associated with rotational
motion is about two orders of magnitude larger than that of
translational motion, solely due to the corresponding difference in
spatial scales (mean free path vs. the molecule size). Note that this
difficulty is not related to any interactions, only the fact that the
rotational motion of a gas molecule is in itself less deterministic
than its translational motion.

\subsection{Our model of interactions between polyatomic molecules}

From what is outlined above, the obvious conclusion is -- not only
tracking the angles and angular velocities of molecules, as well as
quantifying their collisions in a detailed fashion, is complicated
technically, it is also largely meaningless from the standpoint of
physics. Even if one develops a kinetic model which tracks the angular
motions and interactions of polyatomic molecules exactly and
deterministically, it is not clear what benefit it would provide if
applied to realistic gases at normal conditions. Thus, a reasonable
way to proceed from here is to resort to a more general, stochastic
model of rotational collisions.

Here, we will take a semi-empirical approach to the kinetic
description of polyatomic molecular interactions. What this means is
that we will formulate a general mathematical model of the interaction
of two polyatomic gas molecules with basic properties, but, at the
same time, will not elaborate on the specific details of the
interaction (as that may be quite complicated). This will lead to a
correct mathematical form of the kinetic equation, where the
rotational collision integral possesses all necessary general
properties. We will neglect interactions of three or more molecules at
once -- for most gases at normal temperature and pressure, such
interactions are not statistically significant.

In the context of our model, the interaction of a gas molecule with
another gas molecule can be separated into two distinct effects.
\begin{enumerate}
\item The first effect is the deterministic interaction via an
  intermolecular potential, where the potential depends only on the
  translational coordinates, and whose gradient accelerates only the
  linear velocities. The rotational coordinates and angular velocities
  are unaffected by the potential interaction.
\item The second effect is the stochastic collisional interaction,
  which generally depends on all degrees of freedom, and affects both
  linear and angular velocities. This effect is responsible for the
  exchange of the momentum and energy between the linear and angular
  velocities of both involved molecules.
\end{enumerate}
While the first effect is easy to describe by introducing a suitable
intermolecular potential $\phi(r)$, the second effect needs a more
detailed elaboration. Namely, we have to formulate a collision model,
which is general enough to be capable of describing a broad range of
interactions between polyatomic gas molecules, and at the same time is
simple enough to be treated analytically to the extent needed. In
order to accomplish this, we employ the same mathematical mechanism
which we used in our work \cite{Abr17}.

In what follows, we assume that collisional interactions occur
instantaneously -- that is, when a collision occurs, linear and
angular velocities of both colliding molecules are changed
instantaneously to new values. Thus, the description of a collisional
interaction must involve two parts -- {\em when} the interaction
occurs, and {\em how} it occurs. Below, we start with the latter,
followed by the former.

\subsection{The general mechanics of a polyatomic collision}

In what follows, we assume that the total number of degrees of freedom
of each gas molecule is $N$, of which 3 are translational, and $N-3$
are rotational. We denote the translational coordinates via $\BV x$,
rotational via $\BV y$, while their respective linear and angular
velocities are given via $\BV v$ and $\BV w$.  Let the linear and
angular velocities of the two colliding molecules be denoted via $(\BV
v_1,\BV w_1)$ and $(\BV v_2,\BV w_2)$, respectively. We assume that,
whenever the collision occurs, the velocities are changed
instantaneously via a collision mapping $\collmap$ to $(\BV v_1',\BV
w_1')$ and $(\BV v_2',\BV w_2')$, respectively, where the increment
depends only on the differences of the pre-collisional values of the
coordinates and velocities, that is,
\begin{equation}
\label{eq:collision}
(\BV v_1',\BV w_1',\BV v_2',\BV w_2')=\collmap(\BV v_1,\BV w_1,\BV v_2,
\BV w_2)=(\BV v_1+\BV g_{12},\BV w_1+\BV h_{12},\BV v_2+\BV g_{21},\BV
w_2+\BV h_{21}).
\end{equation}
Above, the increments $\BV g_{12}$ and $\BV h_{12}$ depend only on the
differences between the properties of the two molecules, i.e. they are
of the form
\begin{subequations}
\label{eq:gh}
\begin{equation}
\BV g_{12}=\BV g(\BV x_1-\BV x_2,\BV y_1-\BV y_2,\BV v_1-\BV v_2,\BV
w_1-\BV w_2),
\end{equation}
\begin{equation}
\BV h_{12}=\BV h(\BV x_1-\BV x_2,\BV y_1-\BV y_2,\BV v_1-\BV v_2,\BV
w_1-\BV w_2),
\end{equation}
\end{subequations}
with $\BV g_{21}$ and $\BV h_{21}$ obviously resulting from inverting
the signs of all arguments. From the fundamental principles of
physics, the collision must preserve both the linear and angular
momenta, as well as the total kinetic energy:
\begin{subequations}
\label{eq:m_e_conservation}
\begin{equation}
\BV v_1'+\BV v_2'=\BV v_1+\BV v_2,\qquad\BV w_1'+\BV w_2'=\BV w_1+\BV
w_2,
\end{equation}
\begin{equation}
\|\BV v_1'\|^2+\|\BV w_1'\|^2+\|\BV v_2'\|^2+\|\BV w_2'\|^2=
\|\BV v_1\|^2+\|\BV w_1\|^2+\|\BV v_2\|^2+\|\BV w_2\|^2.
\end{equation}
\end{subequations}
The momentum conservation automatically implies
\begin{equation}
\BV g_{21}+\BV g_{12}=\BV 0,\qquad\BV h_{21}+\BV h_{12}=\BV 0,
\end{equation}
which means that $\BV g$ and $\BV h$ are skew-symmetric, that is,
\begin{equation}
\BV g(-\BV z)=-\BV g(\BV z),\qquad\BV h(-\BV z)=-\BV h(\BV z),\qquad
\BV z=(\BV x, \BV y, \BV v, \BV w).
\end{equation}
We also assume that the mapping $\collmap$ from pre-collision to
post-collision velocities, given via \eqref{eq:collision}, is
bijective, that is, there is a unique post-collision state to each
pre-collision state, with a unique inverse. Moreover, the same mapping
transforms the negatives of the post-collision velocities into the
negatives of the pre-collision velocities, that is, the negative of
$\collmap$ is an involution:
\begin{equation}
\label{eq:coll_identity}
-\collmap\circ(-\collmap)=\cI.
\end{equation}
This simply means that if the post-collision velocities are
immediately reversed, then the collision occurs in a reverse manner
and leads to the reversed pre-collision velocities. The latter
identity, in particular, ensures that the Jacobian of $\collmap$ is
unity.

Recall that, in the mechanics of a hard sphere collision, the
collision mapping $\collmap$ itself is also an involution
\citep{Abr17,Gra}.  Here, however, we do not make such an assumption,
since a collision between two polyatomic molecules could be vastly
more complicated than a collision between two hard spheres, and it is
unclear whether such property would hold.

\subsection{The criterion for triggering a collision}

The general mechanics of collision, described above, are
deterministic. However, the conditions, under which those collisions
occur, will be modeled stochastically, as in our recent work
\citep{Abr17}. We assume that, when two molecules are sufficiently
close to each other (that is, $\|\BV x_1-\BV x_2\|$ is sufficiently
small), random collisional interactions may happen with a prescribed
conditional intensity. Such artificial randomness reflects the
presence of the Heisenberg uncertainty associated with the angular
orientation and velocities of the pair of interacting molecules.

Mathematically, the collisions will be triggered via increments of a
Poisson process with conditional intensity $\lambda$
\citep{Papa,DalVer}, where the latter depends on the differences of
coordinates and velocities:
\begin{equation}
\lambda_{12}=\lambda(\BV x_1-\BV x_2,\BV y_1-\BV y_2,\BV v_1-\BV
v_2,\BV w_1-\BV w_2).
\end{equation}
This conditional intensity must satisfy the following properties:
\begin{enumerate}
\item $\lambda$ is invariant with respect to the reordering of the
  molecules, that is
\begin{equation}
\label{eq:lambda_reordering}
\lambda(-\BV x,-\BV y,-\BV v,-\BV w)=\lambda(\BV x,\BV y,\BV v,\BV w).
\end{equation}
\item $\lambda$ is the same for reversed post-collision velocities,
  that is,
\begin{equation}
\label{eq:lambda_reverse}
\lambda(\BV x,\BV y,-\BV v',-\BV w')=\lambda(\BV x,\BV y,\BV v,\BV w),
\end{equation}
since, for the same angles, in the time-backward configuration the
collision must be as likely to occur as in the time-forward
configuration.
\end{enumerate}
In what follows, we will generally presume that $\lambda$ is zero when
the two molecules are spaced apart by a considerable distance, so that
the collisions are not triggered. Once the molecules approach each
other, $\lambda$ becomes nonzero, thus possibly triggering collisions.
The dependence of $\lambda$ on the angular coordinates and velocities
additionally provides for configuring the fine details of a collision
criterion however necessary. For additional convenience, here we
impose the following simplifying assumption on $\lambda$:
\begin{equation}
\label{eq:lambda_post_coll}
\lambda(\BV x,\BV y,\BV v',\BV w')=\lambda(\BV x,\BV y,\BV v,\BV w),
\end{equation}
that is, the intensity of collisions does not change due to a
collision, which, particularly, may result in more than one rotational
collision interaction during a single pass of one molecule by another.
The latter assumption leads, together with
\eqref{eq:lambda_reordering} and \eqref{eq:lambda_reverse}, to
$\lambda$ being fully symmetric under the collisions and reordering of
molecules:
\begin{multline}
\label{eq:lambda_full_symmetry}
\lambda(\BV x,\BV y,\BV v,\BV w)=\lambda(\BV x,\BV y,\BV v',\BV w')
=\lambda(\BV x,\BV y,-\BV v,-\BV w)=\lambda(\BV x,\BV y,-\BV v',-\BV
w')\\=\lambda(-\BV x,-\BV y,\BV v',\BV w')=\lambda(-\BV x,-\BV y,\BV
v,\BV w)=\lambda(-\BV x,-\BV y,-\BV v',-\BV w')=\lambda(-\BV x,-\BV
y,-\BV v,-\BV w).
\end{multline}
This symmetry is markedly different from the setting we previously
used for the random hard sphere collisions in \citep{Abr17}, where the
conditional intensity of the Poisson process was set to zero in a
post-collision state. In the present context, however, the molecules
are repelled via the intermolecular potential $\phi$ (which was absent
in \citep{Abr17}), while the collisions are only used to model
additional interactions emerging from the presence of rotational
degrees of freedom. Thus, we do not find the condition in
\eqref{eq:lambda_post_coll} too restrictive.

\section{The dynamical system of polyatomic molecules}
\label{sec:dyn_sys}

With what is formulated above, the evolution equations for a system of
$K$ polyatomic molecules, whose linear and angular coordinates and
velocities are expressed via $\BV x_i$, $\BV y_i$, $\BV v_i$ and $\BV
w_i$, is given via the following L\'evy-type Feller process
\cite{Cou,GikSko,Fel2,App}:
\begin{subequations}
\label{eq:dyn_sys}
\begin{equation}
\deriv{\BV x_i}t=\BV v_i,\qquad\deriv{\BV y_i}t=\BV w_i,
\end{equation}
\begin{equation}
\dif\BV v_i=\sum_{j\neq i}\left[-\parderiv{}{\BV x_i}\phi(\|\BV x_i
  -\BV x_j\|)\dif t+\int_{\RR>0}\BV g_{ij}\xi M_{ij}(\dif t,\dif\xi)
  \right],
\end{equation}
\begin{equation}
\dif\BV w_i=\sum_{j\neq i}\int_{\RR>0}\BV h_{ij}\xi M_{ij}(\dif t,\dif
\xi).
\end{equation}
\end{subequations}
Above, $\phi(r)$ is the intermolecular potential, and
$M_{ij}(t,\cdot)$ is a Poisson random measure with the intensity
$\lambda_{ij}$, which models rotational collisions between the $i$th
and $j$th molecules. In a similar manner as in
\citep{Abr17,Abr20,Abr22}, we concatenate
\begin{equation}
\BV X=(\BV x_1,\BV y_1,\ldots\BV x_K,\BV y_K),\qquad\BV V=(\BV v_1,\BV
w_1,\ldots,\BV v_K,\BV w_K),
\end{equation}
and denote
\begin{equation}
\Phi(\BV X)=\sum_{i=1}^{K-1}\sum_{j=i+1}^K\phi(\|\BV x_i-\BV x_j\|),
\qquad
\BV G_{ij}=(\BV 0,\ldots,\BV 0,\BV g_{ij},\BV h_{ij},\BV 0,\ldots,\BV
0,\BV g_{ji},\BV h_{ji},\BV 0,\ldots,\BV 0),
\end{equation}
where $\BV g_{ij}$ and $\BV h_{ij}$ occupy the slots corresponding to
$\BV v_i$ and $\BV w_i$, respectively, while $\BV g_{ji}$ and $\BV
h_{ji}$ occupy the slots corresponding to $\BV v_j$ and $\BV w_j$,
respectively, with $i<j$. Then, the dynamical system in
\eqref{eq:dyn_sys} can be written in the form
\begin{equation}
\label{eq:dyn_sys_X}
\dif\begin{pmatrix}\BV X \\ \BV V\end{pmatrix}=\begin{pmatrix}\BV V
\\ -\partial\Phi/\partial\BV X\end{pmatrix}\dif t+\sum_{i=1}^{K-1}
\sum_{j=i+1}^K\int_{\RR>0}\begin{pmatrix}\BV 0 \\ \BV G_{ij}
\end{pmatrix} \xi M_{ij}(\dif t,\dif\xi).
\end{equation}
Obviously, the system above preserves the total energy of the system
along its trajectory,
\begin{equation}
E=\frac 12\|\BV V(t)\|^2+\Phi(\BV X(t))=\text{const},
\end{equation}
which follows from the fact that the collisions preserve the kinetic
energy part of the above expression, and do not affect the
coordinates.

\subsection{The forward Kolmogorov equation and its steady states}

The time derivative of the conditional expectation of a test function
$\psi$ is given via the infinitesimal generator
of~\eqref{eq:dyn_sys_X},
\begin{equation}
\label{eq:generator}
\parderiv{}t\EE[\psi]=\BV V\cdot\parderiv\psi{\BV X}-\parderiv\Phi{\BV
  X}\cdot\parderiv\psi{\BV V}+\sum_{i=1}^{K-1}\sum_{j=i+1}^K
\lambda_{ij}\big(\psi(\collmap_{ij}(\BV V))-\psi(\BV V)\big),
\end{equation}
where $\collmap_{ij}$ is the collision mapping for the $i$th and $j$th
velocities. For details on the infinitesimal generators of L\'evy-type
Feller processes, see \citep{Cou,GikSko,Fel2,App}. The forward
Kolmogorov equation for the density of states $F$ is obtained in the
same manner as in \citep{Abr17}: we multiply \eqref{eq:generator} by
$F$, integrate over $\BV X$ and $\BV V$, change the variable of
integration in the first term of the collision integral from
$\collmap_{ij}(\BV V)$ to $\BV V$, and then strip the integral
together with $\psi$, while making use of
\eqref{eq:lambda_full_symmetry}, and the fact that the Jacobian of
$\collmap_{ij}$ is unity. The resulting forward Kolmogorov equation is
given via
\begin{equation}
\label{eq:kolmogorov}
\parderiv Ft+\BV V\cdot\parderiv F{\BV X}=\parderiv\Phi{\BV X}\cdot
\parderiv F{\BV V}+\sum_{i=1}^{K-1}\sum_{j=i+1}^K\lambda_{ij}
\big(F(\collmap_{ij}^{-1}(\BV V))-F(\BV V)\big).
\end{equation}
It is easy to find some steady states of \eqref{eq:kolmogorov} if we
require that $F$ is invariant under collisions, that is
\begin{equation}
\label{eq:F_inv}
F(\collmap_{ij}(\BV V))=F(\BV V),
\end{equation}
for all pairs of molecules. For such $F$, the forward Kolmogorov
equation \eqref{eq:kolmogorov} becomes
\begin{equation}
\label{eq:kolmogorov_no_collisions}
\parderiv Ft+\BV V\cdot\parderiv F{\BV X}=\parderiv\Phi{\BV X}
\cdot\parderiv F{\BV V}.
\end{equation}
Next, we note that $F=F_0(E)$ satisfies \eqref{eq:F_inv} (since the
energy is invariant under the collisions), and is at the same time the
steady state for \eqref{eq:kolmogorov_no_collisions}, which means that
it is automatically the steady state for the forward Kolmogorov
equation \eqref{eq:kolmogorov}. Among all such states, the canonical
Gibbs state is given via
\begin{equation}
F_G=\frac 1{(2\pi\theta_0)^{KN/2}Z_K\Omega^K}\exp\left(-\frac{\|\BV
  V\|^2+2 \Phi(\BV X)}{2\theta_0}\right),\quad Z_K=\int e^{-\Phi/
  \theta_0}\dif\BV x_1\ldots\dif\BV x_K,
\end{equation}
where $\Omega$ is the full solid angle of the domain of rotational
coordinates, and $\theta_0$ is the equilibrium kinetic temperature of
the system of molecules. Here, observe that, thanks
to~\eqref{eq:lambda_post_coll} and~\eqref{eq:lambda_full_symmetry},
the presence of stochastic rotational collisions does not affect the
structure of the steady state; this is to the contrary of what we had
previously in \citep{Abr17}, where the asymmetry of the intensity of
collisions created the ``potential well'' instead.

A solution of the forward Kolmogorov equation \eqref{eq:kolmogorov}
possesses the following entropy inequalities:
\begin{equation}
\label{eq:entropy}
-\parderiv{}t\int F\ln F\dif\BV X\dif\BV V\geq 0,\qquad\parderiv{
}t\int F\ln\left(\frac F{F_0}\right)\dif\BV X\dif\BV V\leq 0,
\end{equation}
where $F_0$ is a steady state of \eqref{eq:kolmogorov}. Above, the
first quantity is the Shannon entropy~\citep{Shan}, while the second
one is the Kullback--Leibler entropy~\citep{KulLei}. The proof of the
relations in~\eqref{eq:entropy} is given in
Appendix~\ref{sec:entropy_inequality}.

\subsection{The structure of a two-molecule marginal distribution of
the Gibbs state}

Let us integrate $F_G$ over all molecules but the first two. This
integration decomposes into the product of integrals over the
velocities and coordinates separately:
\begin{equation}
F_G^{(2)}=\frac 1{(2\pi\theta_0)^N\Omega^2}e^{-(\|\BV v_1\|^2+\|\BV
  w_1\|^2+\| \BV v_2\|^2+\|\BV w_2\|^2)/2\theta_0}\frac 1{Z_K}\int
e^{-\Phi/\theta_0}\dif\BV x_3\ldots\dif\BV x_K.
\end{equation}
Next, note that the single-particle marginal distribution $f_G$ is
given via
\begin{equation}
f_G(\BV v,\BV w)=\frac 1{(2\pi\theta_0)^{N/2}V\Omega}e^{-(\|\BV
  v\|^2+\|\BV w\|^2)/2\theta_0},
\end{equation}
where $V$ is the volume of the domain of translational coordinates.
The reason for the form above is that, spatially, $F_G$ depends only
on the differences of the translational coordinates, and thus the
integration over all molecules but one removes the spatial dependence
completely. Thus, in terms of $f_G$, $F_G^{(2)}$ can be written in the
form
\begin{equation}
F_G^{(2)}=f_G(\BV v_1,\BV w_1)f_G(\BV v_2,\BV w_2)\frac{V^2}{Z_K}\int
e^{-\Phi/\theta_0}\dif\BV x_3\ldots\dif\BV x_K.
\end{equation}
Next, let us look at the factor which multiplies the product of
$f_G$'s. It can be written in the form
\begin{equation}
\frac{V^2}{Z_K}\int e^{-\Phi/\theta_0}\dif\BV x_3\ldots\dif\BV
x_K=\frac K{K-1}e^{-\phi(\|\BV x_1-\BV x_2\|)/\theta_0}
Y_K(\theta_0,\|\BV x_1 -\BV x_2\|),
\end{equation}
where $Y_K(\theta_0,r)$ is the pair cavity distribution function for
$K$ molecules \citep{Bou86,Bou06}:
\begin{multline}
Y_K(\theta_0,\|\BV x_1-\BV x_2\|)\\=\frac{K-1}K\frac{V^2}{Z_K}\int
\prod_{i=3}^K e^{-(\phi(\| \BV x_1-\BV x_i\|)+\phi(\|\BV x_2-\BV x_i
  \|) )/\theta_0}\prod_{ j=i+1}^K e^{-\phi(\|\BV x_i-\BV x_j\|)
  /\theta_0}\dif\BV x_3 \ldots\dif\BV x_K.
\end{multline}
Thus, with help of $Y_K(\theta_0,r)$, $F_G^{(2)}$ is given via
\begin{equation}
\label{eq:F_G_2}
F_G^{(2)}=\frac K{K-1}e^{-\phi(\|\BV x_1- \BV x_2\|)/\theta_0} Y_K(
\theta_0,\|\BV x_1-\BV x_2\|)f_G(\BV v_1,\BV w_1)f_G(\BV v_2,\BV w_2).
\end{equation}

\section{The equation for the distribution of a single molecule}
\label{sec:boltzmann-vlasov}

Let us integrate the forward Kolmogorov equation \eqref{eq:kolmogorov}
over all molecules but the first one, and, for convenience, denote
$\BV z_i=(\BV x_i,\BV y_i,\BV v_i,\BV w_i)$:
\begin{multline}
\label{eq:BBGKY}
\parderiv ft+\BV v\cdot\parderiv f{\BV x}+\BV w\cdot\parderiv f{\BV
  y}=\sum_{i=2}^K\int\bigg[\parderiv{}{\BV x}\phi(\|\BV x-\BV x_i\|)
  \cdot\parderiv{F^{(2)}_{1,i}(\BV z,\BV z_i)}{\BV v}\\+\lambda_{1i}
  \big(F^{(2)}_{1,i}(\collmap^{-1}(\BV z,\BV z_i))-F^{(2)}_{1,i}(\BV z,
  \BV z_i) \big)\bigg]\dif\BV z_i.
\end{multline}
This equation constitutes the first iteration of the
Bogoliubov--Born--Green--Kirkwood--Yvon \citep{Bog,BorGre,Kir}
hierarchy (BBGKY). In order to obtain a closure for the right-hand
side of \eqref{eq:BBGKY} in terms of $f$, as previously in
\citep{Abr17,Abr22,Abr23}, we assume that all pair distributions
$F^{(2)}_{1,i}$ are identical, which leads to
\begin{multline}
\parderiv ft+\BV v\cdot\parderiv f{\BV x}+\BV w\cdot\parderiv f{\BV
  y}=(K-1)\int\bigg\{\parderiv{}{\BV x}\phi(\|\BV x-\BV x_2\|)
\cdot\parderiv{F^{(2)}(\BV z,\BV z_2)}{\BV v}\\+\lambda_{12}\big[
  F^{(2)}(\collmap^{-1}(\BV z,\BV z_2))-F^{(2)}(\BV z,\BV z_2)\big]
\bigg\}\dif\BV z_2.
\end{multline}
This is a standard assumption in kinetic theory \citep{CerIllPul},
whose purpose is a formal reduction of the multimolecular dynamics to
those of a single molecule, and, therefore, we use it here for the
lack of a better option. However, one must understand that, in
practice, such an assumption can only hold for a relatively small
``parcel'' of gas (confined, for example, to a periodic cube whose
size is not much larger than the mean free path), otherwise there
would be no reason for different pairs of molecules to have identical
distributions. As a result, a kinetic interaction of such a parcel
with surroundings is not accounted for in the context of such a
formalism, which below leads to an empirical closure of the moment
hierarchy in an attempt to accommodate such unaccounted interactions.

Next, we need to approximate $F^{(2)}$ in terms of $f$; as we have
done previously in \citep{Abr17,Abr22}, we assume that $F^{(2)}$ has
the same form as does $F_G^{(2)}$ in \eqref{eq:F_G_2}:
\begin{equation}
F^{(2)}(\BV z,\BV z_2)=\frac K{K-1}e^{-\phi(\|\BV x-\BV x_2\|)/\theta}
Y_K(\theta,\|\BV x-\BV x_2\|)f(\BV z)f(\BV z_2).
\end{equation}
Above, $\theta$ is no longer an equilibrium temperature; instead, it
is now the average kinetic energy of the given molecule (and thus an
appropriate moment of $f$ itself), which endows it with a dependence
on $\BV x$. Thus, for symmetry, in the closure above we
compute~$\theta$ at the midpoint between $\BV x$ and $\BV x_2$:
$\theta=\theta((\BV x+\BV x_2)/2)$. In addition, we renormalize $f$ so
that it is the mass density, $f\to Kmf$, where $m$ is the mass of a
single molecule. Applying the closure and the rescaling, we arrive at
the following closed equation:
\begin{multline}
\label{eq:transport2}
\parderiv ft+\BV v\cdot\parderiv f{\BV x}+\BV w\cdot\parderiv f{\BV y}
= \frac 1m\int e^{-\phi(\|\BV x-\BV x_2\|)/\theta((\BV x+\BV x_2)/2)}
Y_K(\theta((\BV x+\BV x_2)/2),\|\BV x- \BV x_2\|)\\\bigg(\parderiv{}{
  \BV x}\phi(\|\BV x-\BV x_2\|)\cdot \parderiv{f(\BV z)}{\BV v}f(\BV
z_2)+\lambda(\BV z-\BV z_2)\big(f(\BV z'')f(\BV z_2'')-f(\BV z)f(\BV
z_2)\big) \bigg)\dif\BV z_2,
\end{multline}
where $\BV z''$ and $\BV z_2''$ refer to the inverse of the collision
mapping, $(\BV z'',\BV z_2'')=\collmap^{-1}(\BV z,\BV z_2)$. Above,
the kinetic temperature $\theta$ is given as follows: let us introduce
the average $\langle\psi\rangle$ of a function $\psi(\BV x,\BV y,\BV
v,\BV w)$ via
\begin{equation}
\label{eq:moment}
\langle\psi\rangle(t,\BV x)=\int\psi(\BV x,\BV y,\BV v,\BV w) f(t,\BV
x,\BV y,\BV v,\BV w)\dif\BV y\dif\BV v\dif\BV w.
\end{equation}
Then, we define the density $\rho$, linear and angular velocities $\BV
u$ and $\BV u_{\BV w}$, respectively, and the kinetic temperature
$\theta$ via the following velocity moments:
\begin{equation}
\label{eq:rho_u_p}
\rho=\langle 1\rangle,\qquad\rho\BV u=\langle\BV v\rangle,\qquad \rho\BV
u_{\BV w}=\langle\BV w\rangle,\qquad\rho\theta=\frac 1N\langle\|\BV v-
\BV u\|^2 +\|\BV w-\BV u_{\BV w}\|^2\rangle.
\end{equation}
Above, the product $\rho\theta$ can be referred to as the ``kinetic''
pressure, defined through the equation of state for an ideal gas -- as
opposed to the ``true'' van der Waals' pressure, to be found below.
With the above notations, we can separate the integrals in
\eqref{eq:transport2} as follows:
\begin{multline}
\label{eq:transport}
\parderiv ft+\BV v\cdot\parderiv f{\BV x}+\BV w\cdot\parderiv f{\BV y}
\\= \parderiv{f(\BV z)}{\BV v}\cdot\frac 1m\int e^{-\frac{\phi(\|\BV
    x-\BV x_2\|)}{\theta((\BV x+\BV x_2)/2)}} Y_K(\theta((\BV x+\BV
x_2)/2),\|\BV x- \BV x_2\|)\parderiv{}{ \BV x}\phi(\|\BV x-\BV
x_2\|)\rho(\BV x_2)\dif\BV x_2\\+\frac 1m\int e^{-\frac{\phi(\|\BV x-\BV
    x_2\|)}{\theta((\BV x+\BV x_2)/2)}} Y_K(\theta((\BV x+\BV
x_2)/2),\|\BV x- \BV x_2\|)\lambda(\BV z-\BV z_2)\big(f(\BV
z'')f(\BV z_2'')-f(\BV z)f(\BV z_2)\big)\dif\BV z_2.
\end{multline}

\subsection{Hydrodynamic limit}

For most practical situations, the size of each molecule is much
smaller than the size of the domain. If so, the transport equation for
$f$ can be simplified further by computing what is known as the
``hydrodynamic limit'' -- that is, an appropriate limit where the size
of a molecule becomes infinitely small in comparison with the size of
the domain. To this end, we introduce the constant parameter $\sigma$,
which is to be the ``diameter'' of a molecule, and, with its help,
rescale the distance between the molecules in all of the properties of
molecular interactions:
\begin{subequations}
\label{eq:rescaling}
\begin{equation}
\phi(r)\to\phi(r/\sigma),\qquad Y_K(\theta,r)\to Y_K(\theta,r/\sigma),
\qquad\lambda(\BV x,\BV y,\BV v,\BV w)\to\lambda(\BV x/\sigma,\BV
y,\BV v,\BV w),
\end{equation}
\begin{equation}
\BV g(\BV x,\BV y,\BV v,\BV w)\to\BV g(\BV x/\sigma,\BV y,\BV v,\BV
w),\qquad\BV h(\BV x,\BV y,\BV v,\BV w)\to\BV h(\BV x/\sigma,\BV y,\BV
v,\BV w).
\end{equation}
\end{subequations}
Then, as $\sigma\to 0$, $m\to 0$, and $K\to\infty$, with $m/\sigma^3$
and $Km$ being constants (so that the density of a molecule and the
total mass of the system would be fixed), the
equation~\eqref{eq:transport} is transformed as follows:
\begin{equation}
\label{eq:boltzmann-vlasov}
\parderiv ft+\BV v\cdot\parderiv f{\BV x}+\BV w\cdot\parderiv f{\BV y}
=\frac 1\rho\parderiv{\bar\phi}{\BV x}\cdot\parderiv f{\BV
  v}+\coll(f).
\end{equation}
Above, the potential forcing and the collision integral in the
right-hand side are given, respectively, via
\begin{subequations}
\label{eq:forcing_collision}
\begin{equation}
\label{eq:bphi}
\bar\phi=\frac{2\pi}3\frac{\sigma^3}m\rho^2(\BV x)\theta(\BV x)
\int_0^\infty\left(1-e^{-\phi(r)/\theta(\BV x)}\right)\parderiv{}r
\big(r^3Y(\theta(\BV x),r)\big)\dif r,
\end{equation}
\begin{multline}
\label{eq:collision_integral}
\coll(f)=\frac{\sigma^3}m\int\alpha(\BV x,\BV r,\BV y-\BV y_2,\BV
v-\BV v_2,\BV w-\BV w_2)\\\big(f(\BV x,\BV y,\BV v'',\BV w'')f(\BV
x,\BV y_2,\BV v_2'',\BV w_2'')-f(\BV x,\BV y,\BV v,\BV w)f(\BV x,\BV
y_2,\BV v_2,\BV w_2)\big)\dif\BV r\dif\BV y_2\dif\BV v_2\dif\BV w_2,
\end{multline}
\begin{equation}
\label{eq:alpha}
\alpha(\BV x,\BV r,\BV y,\BV v,\BV w)=\lambda(\BV r,\BV y,\BV v,\BV
w)e^{-\phi(\|\BV r\|)/\theta(\BV x)} Y(\theta(\BV x),\|\BV r\|),
\end{equation}
\end{subequations}
where the integration in $\dif\BV r$ occurs over $\RR^3$.  Below, we
refer to \eqref{eq:boltzmann-vlasov} as the {\em Boltzmann--Vlasov
  equation}, because it contains both the deterministic potential (as
in the Vlasov equation~\citep{Vla}), and the stochastic collision
integral (as in the Boltzmann equation \citep{Bol}).

In \eqref{eq:forcing_collision}, $Y(\theta,r)$ denotes the cavity
distribution function for infinitely many molecules, and the
increments in the collision mapping $\collmap$ are computed as
\begin{subequations}
\label{eq:gh_rescaled}
\begin{equation}
\BV g_{12}=\BV g(\BV r,\BV y_1-\BV y_2,\BV v_1-\BV v_2,\BV w_1-\BV
w_2),
\end{equation}
\begin{equation}
\BV h_{12}=\BV h(\BV r,\BV y_1-\BV y_2,\BV v_1-\BV v_2,\BV w_1-\BV
w_2),
\end{equation}
\end{subequations}
The derivation of the forcing and collision terms in
\eqref{eq:forcing_collision} is presented in
Appendix~\ref{app:hydrodynamic_limit}. Since it is assumed that the
volume of the domain is constant, we can see that the packing fraction
(that is, the total volume of molecules divided by the volume of the
domain) also approaches a constant, and, in particular, does not
vanish in the hydrodynamic limit:
\begin{equation}
K\sigma^3\cdot\frac 1V=(Km)\frac{\sigma^3}m\frac 1V\sim \text{const}.
\end{equation}
Generally, the cavity distribution function $Y(\theta,r)$ also depends
on the packing fraction.

\subsection{Steady state of the collision integral}

Observe that, irrespectively of $\alpha$, the collision integral in
\eqref{eq:collision_integral} is guaranteed to be zero when
\begin{equation}
f(\BV x,\BV y,\BV v'',\BV w'')f(\BV x,\BV y_2,\BV v_2'',\BV w_2'')
=f(\BV x,\BV y,\BV v,\BV w)f(\BV x,\BV y_2,\BV v_2,\BV w_2),
\end{equation}
or, if we take a logarithm on both sides,
\begin{equation}
\ln f(\BV x,\BV y,\BV v'',\BV w'')+\ln f(\BV x,\BV y_2,\BV v_2'',\BV
w_2'')=\ln f(\BV x,\BV y,\BV v,\BV w)+\ln f(\BV x,\BV y_2,\BV v_2,\BV
w_2).
\end{equation}
Taking into account the momentum and kinetic energy conservation laws
of collision \eqref{eq:m_e_conservation}, we conclude that the above
identity holds under the same conditions as in the usual Boltzmann
equation \citep{Bol,ChaCow,CerIllPul,HirCurBir}, that is,
\begin{equation}
\ln f=a_1+\BV a_2\cdot\BV v+\BV a_3\cdot\BV w+a_4(\|\BV v\|^2+\|\BV
w\|^2),
\end{equation}
for arbitrary $a_1$, $\BV a_2$, $\BV a_3$ and $a_4$. As a result, the
Maxwell--Boltzmann equilibrium state
\begin{equation}
f_{MB}=\frac\rho{(2\pi\theta)^{N/2}}\exp\left(-\frac{\|\BV v-\BV
    u\|^2+\|\BV w-\BV u_{\BV w}\|^2}{2\theta}\right),
\end{equation}
which has the correct values of the moments in \eqref{eq:rho_u_p}, is
a steady state for the collision integral alone. Additionally, if
$\rho$, $\BV u$, $\BV u_{\BV w}$ and $\theta$ do not depend on $\BV
x$, then $f_{MB}$ becomes a steady state for the whole
Boltzmann--Vlasov equation in~\eqref{eq:boltzmann-vlasov}.

\subsection{Moments of the collision integral}

We define a moment of the collision integral via
\begin{multline}
\langle\psi\rangle_\coll(t,\BV x)=\int\psi(\BV x,\BV y,\BV v,\BV w)
\coll(f)\dif\BV y\dif\BV v\dif\BV w=\int\alpha(\BV x,\BV r,\BV y-\BV
y_2,\BV v-\BV v_2,\BV w-\BV w_2)\\\big(\psi(\BV x,\BV y,\BV v',\BV
w')-\psi(\BV x,\BV y,\BV v,\BV w)\big) f(\BV x,\BV y,\BV v,\BV w)f(\BV
x,\BV y_2,\BV v_2,\BV w_2)\dif\BV r\dif\BV y_2\dif\BV v_2\dif\BV
w_2\dif\BV y\dif\BV v\dif\BV w,
\end{multline}
where in the second identity we changed the collision mapping from
backward to forward under the integral. Observe that, due to
\eqref{eq:lambda_full_symmetry}, $\langle\psi\rangle_\coll$ is
invariant under the permutation of the two molecules:
\begin{multline}
\label{eq:coll_permutation}
\langle\psi\rangle_\coll(t,\BV x)=\frac 12\int\alpha(\BV x,\BV r,\BV
y-\BV y_2,\BV v-\BV v_2,\BV w-\BV w_2)\big(\psi(\BV x,\BV y,\BV v',\BV
w')+ \psi(\BV x,\BV y_2,\BV v_2',\BV w_2')\\-\psi(\BV x,\BV y,\BV
v,\BV w)-\psi(\BV x,\BV y_2,\BV v_2,\BV w_2) \big)f(\BV x,\BV y,\BV
v,\BV w)f(\BV x,\BV y_2,\BV v_2,\BV w_2)\dif\BV r\dif\BV y_2\dif\BV
v_2\dif\BV w_2\dif\BV y\dif\BV v\dif\BV w.
\end{multline}
Due to the momentum and kinetic energy conservation laws of the
collision mechanics, we can see that the collision moments are zero
for $\psi=1, \BV v,\BV w$, $(\|\BV v\|^2+\|\BV w\|^2)$, and,
subsequently, $\ln f_{MB}$. Additionally, we show in
Appendix~\ref{app:h-theorem} that
\begin{equation}
\label{eq:lnf_inequality}
\langle\ln f\rangle_\coll\leq 0.
\end{equation}
Then, it is easy to verify that the following inequalities hold for
the Shannon entropy $\langle-\ln f\rangle$ and the Kullback--Leibler
entropy $\langle\ln(f/f_{MB})\rangle$:
\begin{equation}
\parderiv{}t\langle-\ln f\rangle+\nabla_{\BV x}\cdot\langle-\BV v\ln f
\rangle\geq 0,\qquad\parderiv{}t\langle\ln(f/f_{MB})\rangle+\nabla_{
  \BV x}\cdot\langle\BV v\ln(f/f_{MB})\rangle\leq 0.
\end{equation}
The first inequality constitutes Boltzmann's $H$-theorem for
\eqref{eq:boltzmann-vlasov}.

\subsection{Velocity moment equations}

To obtain the transport equation for a velocity moment of the
form~\eqref{eq:moment}, we integrate the Boltzmann--Vlasov equation in
\eqref{eq:boltzmann-vlasov} against $\psi(\BV v,\BV w)$, and assume
that there are no boundary effects when the forcing is integrated by
parts:
\begin{equation}
\parderiv{\langle\psi\rangle}t+\nabla\cdot\langle\psi\BV v\rangle=-
\frac 1\rho\nabla\bar\phi\cdot\langle\nabla_{\BV v}\psi\rangle+
\langle\psi\rangle_\coll.
\end{equation}
Above, ``$\nabla$'' without a subscript denotes the differentiation in
$\BV x$.  From \eqref{eq:m_e_conservation} and
\eqref{eq:coll_permutation}, it automatically follows that, for
$\psi=1, \BV v, \BV w$ and $(\|\BV v\|^2+\|\BV w\|^2)$, the collision
integral disappears. In particular, the transport equation for the
density $\rho$ is given via
\begin{equation}
\label{eq:mass}
\parderiv{\langle 1\rangle}t+\nabla\cdot\langle\BV v\rangle=0,\qquad
\text{or}\qquad\parderiv\rho t+\nabla\cdot(\rho\BV u)=0,
\end{equation}
where we recall the notations in \eqref{eq:rho_u_p}. For the momentum
transport equation, we write
\begin{equation}
\parderiv{\langle\BV v\rangle}t+\nabla\cdot\langle\BV v^2\rangle=
-\nabla\bar\phi,\qquad\parderiv{\langle\BV w\rangle}t+ \nabla
\cdot\langle\BV w\BV v^T\rangle=\BV 0,
\end{equation}
where we adopt the convention that the tensor contraction occurs over
the trailing (or column) index. Introducing the stresses
\begin{equation}
\BM S=\langle(\BV v-\BV u)^2\rangle-\rho\theta\BM I,\qquad \langle(\BV v-\BV
u)(\BV w-\BV u_{\BV w})^T\rangle=\BM S_{\BV v\BV w},
\end{equation}
we write the equations for the linear and angular momenta above as
\begin{equation}
\label{eq:momentum}
\parderiv{(\rho\BV u)}t+\nabla\cdot\big(\rho(\BV u^2+\theta\BM I)+\BM
S\big)=-\nabla \bar\phi,\qquad\parderiv{(\rho\BV u_{\BV w})}t+\nabla
\cdot(\rho\BV u_{ \BV w}\BV u^T+\BM S_{\BV w\BV v})=\BV 0.
\end{equation}
For the energy transport equation, we write
\begin{equation}
\parderiv{}t\langle\|\BV v\|^2+\|\BV w\|^2\rangle+\nabla\cdot\langle(
\|\BV v\|^2+\|\BV w\|^2)\BV v\rangle=\frac 1\rho\langle(\|\BV v\|^2+\|
\BV w\|^2)\nabla_{\BV v}\cdot(f\nabla_{\BV x}\bar\phi) \rangle.
\end{equation}
Introducing the heat flux
\begin{equation}
\BV q=\frac 12\langle(\|\BV v-\BV u\|^2+\|\BV w-\BV u_{\BV w} \|^2)
(\BV v-\BV u)\rangle,
\end{equation}
we observe, via simple manipulations, that
\begin{subequations}
\begin{equation}
\langle\|\BV v\|^2+\|\BV w\|^2\rangle=\rho(\|\BV u\|^2+\|\BV u_{\BV w}
\|^2+N\theta),
\end{equation}
\begin{equation}
\langle(\|\BV v\|^2+\|\BV w\|^2)\BV v\rangle=\rho\big(\|\BV u\|^2+\|
\BV u_{\BV w}\|^2+(N+2)\theta\big)\BV u+2(\BM S\BV u+\BM S_{\BV v\BV
  w} \BV u_{\BV w}+\BV q),
\end{equation}
\begin{equation}
\frac 1\rho\langle(\|\BV v\|^2+\|\BV w\|^2)\nabla_{\BV v}\cdot(f
\nabla\bar\phi)\rangle=-2\BV u\cdot\nabla\bar\phi,
\end{equation}
\end{subequations}
which leads to
\begin{multline}
\parderiv{}t\big(\rho(\|\BV u\|^2+\|\BV u_{\BV w}\|^2+N\theta)\big)+
\nabla\cdot\big[\rho\big(\|\BV u\|^2+\|\BV u_{\BV w}\|^2+(N+2)\theta
  \big)\BV u\\+2(\BM S\BV u+\BM S_{\BV w\BV v}\BV u_{\BV w}+\BV q)
  \big]=-2\BV u\cdot\nabla\bar\phi.
\end{multline}
In order to obtain the equation for the kinetic pressure $\rho\theta$,
we express the time derivatives via
\begin{subequations}
\begin{multline}
\parderiv{(\rho\|\BV u\|^2)}t=2\BV u\cdot\parderiv{(\rho\BV u)}t-\|\BV
u\|^2\parderiv\rho t=\|\BV u\|^2\nabla\cdot(\rho\BV u)\\-2\BV u^T
\nabla\cdot(\rho\BV u^2+(\rho\theta+\bar\phi)\BM I+\BM S)=-\nabla\cdot
(\rho\|\BV u \|^2\BV u)-2\BV u\cdot\nabla(\rho\theta+\bar\phi)-2\BV
u^T\nabla\cdot\BM S,
\end{multline}
\begin{multline}
\parderiv{(\rho\|\BV u_{\BV w}\|^2)}t=2\BV u_{\BV w}\cdot\parderiv{
  (\rho\BV u_{\BV w})}t-\|\BV u_{\BV w}\|^2\parderiv\rho t=\|\BV u_{
  \BV w}\|^2\nabla\cdot(\rho\BV u)\\-2\BV u_{\BV w}^T\nabla \cdot(\rho
\BV u_{\BV w}\BV u+\BM S_{\BV w\BV v})=-\nabla\cdot(\rho\|\BV u_{\BV
  w}\|^2\BV u)-2\BV u_{\BV w}^T\nabla\cdot\BM S_{\BV w\BV v},
\end{multline}
\end{subequations}
and subtract from the energy transport equation, which, upon division
by $N$, retains the time derivative for the kinetic pressure only:
\begin{equation}
\label{eq:pressure2}
\parderiv{(\rho\theta)}t+\nabla\cdot(\rho\theta\BV u)+\frac 2N (\rho
\theta\nabla\cdot\BV u+\BM S:\nabla\BV u+\BM S_{\BV v\BV w}:\nabla\BV
u_{\BV w}+\nabla\cdot\BV q)=0.
\end{equation}

\section{A closure for the heat flux based on a prescribed specific heat
  capacity}
\label{sec:closure}

The Boltzmann--Vlasov equation \eqref{eq:boltzmann-vlasov}, and,
subsequently, the transport equations for the density \eqref{eq:mass},
momentum \eqref{eq:momentum} and pressure \eqref{eq:pressure2} are
derived from the multimolecular forward Kolmogorov equation
\eqref{eq:kolmogorov}, under the assumption that the latter represents
a closed system of molecules, which has no interaction with any
outside effects. Moreover, all molecules are regarded to be
statistically identical and thus any molecule is equally likely to
collide with any other molecule, which of course, cannot be true for
domains whose size exceeds the length of the mean free path by many
orders of magnitude. As a result, the Boltzmann--Vlasov equation
practically describes the behavior of a small parcel of gas, which
does not interact with its surroundings at all.

Subsequently, its direct and naive closure, which solely relies upon,
broadly speaking, hypoelliptic properties of the collision integral,
invariably leads to the compressible Enskog--Euler equations
\citep{Abr17,Lac}, where the stress and heat flux are suppressed
toward zero by the collision damping, which emerges from the
Chapman--Enskog expansion \citep{ChaCow,Gra,HirCurBir}. The
Enskog--Euler equations describe a thermodynamic process, where a
parcel of gas, following its trajectory, does not exchange the heat
energy with the neighboring parcels at all. This is a good
approximation for the behavior of a gas at short spatial and temporal
scales, such as shock transitions at supersonic Mach numbers, or
acoustic waves.

However, at longer temporal and spatial scales, the parcel of gas
exchanges the heat energy with neighboring parcels, which results in a
qualitatively different behavior -- such as, for example,
incompressible flow, which has a tendency to manifest at subsonic Mach
numbers. Furthermore, in our recent works \citep{Abr22,Abr23}, we
demonstrated that turbulent behavior in a monatomic gas is observed
for a constant-pressure (or inertial) flow. It is, however, common for
the exchange of momentum to be negligible in practical situations,
since high Reynolds number flows are ubiquitous in nature at
macroscopic scales.

Therefore, in order to arrive at a suitable closure, which
realistically describes a turbulent flow regime, we assume that, while
the stress is negligible as a result of the high Reynolds number
condition, the heat flux is not (even despite the presence of the
collision damping in \eqref{eq:boltzmann-vlasov}), since the gas
exchanges the heat energy with its surroundings. To this end, we
remove the terms with stresses $\BM S$ and $\BM S_{\BV v\BV w}$ from
\eqref{eq:momentum} and \eqref{eq:pressure2} to signify the high
Reynolds number flow, but retain the divergence of the heat flux
$\nabla\cdot\BV q$ in the pressure equation \eqref{eq:pressure2}. Once
the stresses $\BM S$ and $\BM S_{\BV v\BV w}$ are removed, the
pressure equation \eqref{eq:pressure2} becomes decoupled from $\BV
u_{\BV w}$ and $\BM S_{\BV v\BV w}$, and we can discard the equation
for the angular momentum in \eqref{eq:momentum} from
consideration. For reasons which will become clear below, we also
recall the formula for the dimensionless specific heat capacity $c_v$
of an ideal gas at a constant volume,
\begin{equation}
c_v=\frac N2,
\end{equation}
and express $N$ via $c_v$ in the pressure equation
\eqref{eq:pressure2}.  In the formula for $c_v$ above, it is assumed
that the kinetic energy of the gas is distributed uniformly across all
of its degrees of freedom, both translational and rotational.

The above manipulations lead to the following system of transport
equations for the density \eqref{eq:mass}, momentum
\eqref{eq:momentum} and kinetic pressure \eqref{eq:pressure2} at a
high Reynolds number:
\begin{subequations}
\label{eq:moment_equations}
\begin{equation}
\label{eq:mass_mom}
\parderiv\rho t+\nabla\cdot(\rho\BV u)=0,\qquad\parderiv{(\rho\BV u)}t
+\nabla\cdot(\rho\BV u^2)+\nabla(\rho\theta+\bar\phi)=\BV 0,
\end{equation}
\begin{equation}
\label{eq:p2}
\parderiv{(\rho\theta)}t+\nabla\cdot(\rho\theta\BV u)+\frac
1{c_v}(\rho\theta\nabla\cdot\BV u+\nabla \cdot\BV q)=0.
\end{equation}
\end{subequations}
Above, observe that the expression $\rho\theta+\bar\phi$ has the
meaning of the ``true'', or van der Waals, pressure, that is, the
quantity whose gradient constitutes the forcing in the momentum
equation of the gas and thereby accelerates its flow.

Although the transport equations in \eqref{eq:moment_equations} no
longer contain the stress, the divergence of the heat flux
$\nabla\cdot\BV q$ is still present in the pressure equation
\eqref{eq:p2} due to yet unspecified properties of the heat exchange
with the surrounding gas. Thus, we need a closure for the divergence
of the heat flux.

Here, we show that the closure for $\nabla\cdot\BV q$ can be achieved
if one requires the flow to have a prescribed constant specific heat
capacity $c$. Of course, we have to understand that one cannot simply
set the specific heat capacity to an arbitrary constant value and
expect meaningful results; yet, it is known that, for certain values
of $c$, valid thermodynamic processes do indeed exist. For the
purposes of the derivation, however, we will assume that $c$ is an
arbitrary constant parameter.

First, we multiply the whole kinetic pressure equation \eqref{eq:p2}
by a constant $c$:
\begin{equation}
\label{eq:pc}
\parderiv{(c\rho\theta)}t+\nabla\cdot(c\rho\theta\BV u)+\frac c{c_v}(
\rho\theta\nabla\cdot\BV u+\nabla\cdot\BV q)=0.
\end{equation}
Next, we assume that the whole term, multiplied by $c/c_v$, is by
itself equal to $\nabla\cdot\BV q$:
\begin{equation}
\label{eq:closure}
\frac c{c_v}(\rho\theta\nabla \cdot\BV u+\nabla\cdot\BV q)=\nabla
\cdot\BV q,\qquad\text{or}\qquad\nabla\cdot\BV q=\frac c{c_v-c}\,\rho
\theta \nabla \cdot\BV u.
\end{equation}
If so, \eqref{eq:pc} becomes
\begin{equation}
\parderiv{(c\rho\theta)}t+\nabla\cdot(c\rho\theta\BV u+\BV q)=0,
\qquad\text{or} \qquad\parderiv{}t\int_V c\rho\theta\dif V=\oint_S
(c\rho\theta\BV u+\BV q)\cdot\BV n\dif S,
\end{equation}
where $V$ is the volume enclosed by a surface $S$, and we used Gauss'
theorem with the inward unit normal vector $\BV n$. It is clear that
the above equation describes a process with the constant specific heat
capacity equal to $c$; indeed, the rate of increase of the heat energy
in the volume $V$ is the sum of the inward advective and thermal heat
fluxes.

If we assume that, for a prescribed constant specific heat capacity
$c$, the corresponding thermodynamic process indeed exists, then the
relation in \eqref{eq:closure} can be used for the heat flux closure.
Expressing $\nabla\cdot\BV q$ via $\rho\theta\nabla\cdot\BV u$ by
means of \eqref{eq:closure} and substituting it into the kinetic
pressure equation \eqref{eq:p2}, we obtain a closed equation for the
kinetic pressure:
\begin{equation}
\label{eq:p3}
\parderiv{(\rho\theta)}t+\nabla\cdot(\rho\theta\BV u)+\frac{\rho\theta
}{c_v-c}\nabla\cdot\BV u=0.
\end{equation}
The system of equations, consisting of the mass and momentum transport
in \eqref{eq:mass_mom}, and the kinetic pressure transport in
\eqref{eq:p3}, preserves the generalized entropy $\cS$ of the form
\begin{equation}
\label{eq:S}
\cS(c)=\rho\theta^{c-c_v}
\end{equation}
along the stream lines. Indeed, differentiating $\cS$ in time, and
replacing the time derivatives of $\rho$ and $\rho\theta$ with their
respective advection terms from \eqref{eq:mass_mom} and \eqref{eq:p3},
we obtain
\begin{equation}
\label{eq:S_transport}
\parderiv\cS t+\BV u\cdot\nabla\cS=0.
\end{equation}
For computational purposes, it is usually desirable to express a
system of transport equations in the form of conservation laws. In the
present setting, the mass and momentum transport equations in
\eqref{eq:mass_mom} indeed have the conservation law form, however,
the pressure transport equation \eqref{eq:p3} does not. Fortunately,
the quantity $\cA(c)$, given via
\begin{equation}
\label{eq:A}
\cA(c)=\theta^{c_v-c},
\end{equation}
has the transport equation in the form of a conservation law:
\begin{equation}
\label{eq:A_transport}
\parderiv\cA t+\nabla\cdot(\cA\BV u)=0.
\end{equation}
The equations in \eqref{eq:mass_mom}, \eqref{eq:A} and
\eqref{eq:A_transport} constitute a closed system of the transport
equations for the high Reynolds number flow of a gas with the specific
heat capacity $c$.

\subsection{Notable special cases}

As we mentioned above, the equations in \eqref{eq:mass_mom},
\eqref{eq:S} and \eqref{eq:S_transport} do not necessarily describe a
valid process for an arbitrary specific heat capacity~$c$.  However,
there are notable special cases for certain values of $c$, which are
empirically known to be valid thermodynamic processes. Below we list
these known special cases.
\begin{itemize}
\item {\bf Adiabatic process.} Sending $c\to 0$ requires
  $\nabla\cdot\BV q\to 0$ in \eqref{eq:closure} (that is, a parcel of
  gas does not exchange heat energy with the surroundings), which
  yields the compressible Enskog--Euler equations \cite{Abr17,Lac},
  with the entropy $\cS$ given via
  \begin{equation}
  \cS(0)=\rho\theta^{-c_v}.
  \end{equation}
  This regime accurately describes processes at transonic and
  supersonic Mach numbers, such as acoustic waves or shock
  transitions.
\item {\bf Process at constant density.} Sending $c\to c_v$ requires
  $\nabla\cdot\BV u\to 0$ in \eqref{eq:closure}, which leads to the
  incompressible Euler equations \cite{Bat,Gols}. The quantity which
  is preserved along the stream lines is naturally the density $\rho$
  of the gas, that is,
  \begin{equation}
  \cS(c_v)=\rho.
  \end{equation}
  This regime describes processes at intermediate subsonic Mach
  numbers, such as the flow around the airfoil of a piston engine
  aircraft at its typical cruising speed. In classical thermodynamics,
  such process is described by Gay-Lussac's law.
\item {\bf Process at constant temperature.} Sending
  $c\to\infty$ in \eqref{eq:closure} leads to the kinetic pressure
  equation in the same form as that for the density, which means that
  the quantity which is preserved along the stream lines is the
  kinetic temperature $\theta$:
  \begin{equation}
  \parderiv\theta t+\BV u\cdot\nabla\theta=0.
  \end{equation}
  In classical thermodynamics, such process is described by Boyle's
  law.
\item {\bf Process at constant kinetic pressure.}  Setting $c=c_p$,
  where the latter is the specific heat capacity at a constant kinetic
  pressure, and recalling Mayer's relation
  \begin{equation}
  \label{eq:Mayer}
  c_p=c_v+1,\quad \text{we find that} \quad \cS(c_v+1)=\rho\theta,
  \end{equation}
  that is, the kinetic pressure is preserved along the stream lines.
  In classical thermodynamics, such process is described by Charles'
  law.
\end{itemize}

\subsection{Suitable thermodynamic process for a turbulent gas flow
at normal conditions}

At this point, we need to choose one of the four thermodynamic
processes above to model a turbulent flow in realistic conditions.
Here, we note that, normally, turbulence is observed roughly in the
same conditions as is convection in the atmosphere under the effect of
gravity. Convection occurs as follows: when a parcel of air, being
initially in hydrostatic balance between the gravity force and the
pressure gradient, increases its temperature, it simultaneously
expands. However, its pressure remains invariant; as an example,
envision a hot air balloon -- clearly, the pressure inside it is the
same as outside. The expansion decreases the density of the parcel and
upsets the hydrostatic balance, that is, the gravity force, being the
product of the density and the gravity acceleration, is no longer
sufficient to counter the pressure gradient. Thus, the warmer air
rises up.

From what is described above, observe that only the process at
constant kinetic pressure is consistent with the phenomenon of
convection; otherwise, either the gas cannot increase its temperature,
or it cannot expand, or, worse yet, an increase in temperature results
in compression of the gas (adiabatic process). Setting $c=c_v+1$ in
\eqref{eq:p3} leads to
\begin{equation}
\label{eq:pressure}
\parderiv{(\rho\theta)}t+\BV u\cdot\nabla(\rho\theta)=0,
\end{equation}
which, together with the mass and momentum transport equations in
\eqref{eq:mass_mom}, constitutes the same transport equations for a
balanced flow as those in \citep{Abr22,Abr23} for a monatomic gas.
What is more surprising, is that the equations in \eqref{eq:mass_mom}
and \eqref{eq:pressure} no longer contain the specific heat capacity
$c_v$, and, therefore, apply to any polyatomic gas, as long as it
possesses only translational and rotational degrees of freedom. For
numerical computations, we note that
\begin{equation}
\cA(c_v+1)=\theta^{-1},
\end{equation}
which yields the conservation laws for the transport of mass, momentum
and inverse kinetic temperature, identical to those for a monatomic
gas \cite{Abr22,Abr23}:
\begin{equation}
\label{eq:balanced_flow}
\parderiv\rho t+\nabla\cdot(\rho\BV u)=0,\quad\parderiv{(\rho\BV u)}t
+\nabla\cdot(\rho\BV u^2)+\nabla(\rho\theta+\bar\phi)=\BV 0,\quad
\parderiv{(\theta^{-1})}t+\nabla\cdot(\theta^{-1}\BV u)=0.
\end{equation}
In what follows, we study the special case of balanced flow
\eqref{eq:balanced_flow}, where the kinetic pressure throughout the
domain is constant, $\rho\theta=p_0$ (also known as inertial flow). In
such a case, the inverse temperature $\theta^{-1}$ becomes a constant
multiple of the density $\rho$, which makes the equation for the
former redundant. The system of equations for such a flow is thus
comprised solely by the mass and momentum transport equations:
\begin{equation}
\label{eq:inertial_flow}
\parderiv\rho t+\nabla\cdot(\rho\BV u)=0,\qquad\parderiv{(\rho\BV u)}t
+\nabla\cdot(\rho\BV u^2)+\nabla\bar\phi=\BV 0.
\end{equation}

\subsection{Hard sphere approximation for the mean field potential}

Following \citep{Abr22,Abr23}, here we approximate the mean field
potential $\bar\phi$ in \eqref{eq:bphi} via a simple formula, which
can be obtained for a hard sphere interatomic potential. Observe that,
for $\phi$ being that of a hard sphere, that is, zero for $r\geq 1$,
and infinite for $r<1$, the integral in \eqref{eq:bphi} becomes the
cavity distribution function itself,
\begin{equation}
\int_0^\infty\left(1-e^{-\phi_\HS(r)/\theta}\right)\parderiv{}r
\big(r^3Y_\HS(\theta,r)\big)\dif r=\int_0^1\parderiv{}r
\big(r^3Y_\HS(\theta,r)\big)\dif r=Y_\HS(\theta,1),
\end{equation}
which leads to
\begin{equation}
\bar\phi_\HS=\frac{2\pi}3\frac{\sigma^3}m\rho^2\theta Y_\HS(\theta,
1)=\frac{4\rho^2\theta}{\rho_\HS}Y_\HS(\theta,1),\qquad
\rho_\HS=\frac{6m}{\pi\sigma^3}.
\end{equation}
Here, observe that, for hard spheres, the potential forcing in the
momentum equation is the same as the one in the Enskog--Euler
equations \citep{Abr17,Lac} -- that is, the hard sphere collision
integral in the Enskog equation produces the same momentum forcing
term as does the hard sphere potential in the Boltzmann--Vlasov
equation \eqref{eq:boltzmann-vlasov}.

In turn, the cavity distribution function $Y_\HS(\theta, 1)$ for hard
spheres is no longer a function of the temperature, and is only a
function of the packing fraction $\rho/\rho_\HS$ \citep{Bou86,Bou06}:
\begin{equation}
Y_\HS(\theta,1)=\exp\left(\frac 52\frac\rho{\rho_\HS}\right)
+o\left(\frac\rho{\rho_\HS}\right).
\end{equation}
In the numerical simulations that follow, the packing fraction
$\rho/\rho_\HS\sim 10^{-3}$; in such a situation, $Y\approx 1$ and
$\bar\phi$ will be approximated below via
\begin{equation}
\label{eq:bphi_HS}
\bar\phi_\HS\approx\frac{4\rho^2\theta}{\rho_\HS},
\end{equation}
which is the same formula as we used previously in
\citep{Abr22,Abr23}.  For the hard sphere mean field potential given
via \eqref{eq:bphi_HS}, the equations in \eqref{eq:balanced_flow}
become
\begin{subequations}
\label{eq:balanced_flow_HS}
\begin{equation}
\parderiv\rho t+\nabla\cdot(\rho\BV u)=0,\qquad \parderiv{
  (\theta^{-1})}t+\nabla\cdot(\theta^{-1}\BV u)=0,
\end{equation}
\begin{equation}
\parderiv{(\rho\BV u)}t+\nabla\cdot(\rho\BV u^2)+\nabla \bigg[
  \rho\theta\bigg(1+\frac{4\rho}{\rho_\HS}\bigg)\bigg]=\BV 0.
\end{equation}
\end{subequations}
The corresponding inertial flow equations in \eqref{eq:inertial_flow}
become
\begin{equation}
\label{eq:inertial_flow_HS}
\parderiv\rho t+\nabla\cdot(\rho\BV u)=0,\qquad\parderiv{(\rho\BV u)}t
+\nabla\cdot(\rho\BV u^2)+\frac{4p_0}{\rho_\HS}\nabla\rho=\BV 0.
\end{equation}
Remarkably, one can verify that the potential forcing above in
\eqref{eq:inertial_flow_HS} acts in the direction of the temperature
gradient, by substituting $\rho=p_0/\theta$ under the gradient of the
potential term in \eqref{eq:inertial_flow_HS}. Thus, the forcing due
to the intermolecular potential likely creates the thermal creep flow
at constant pressure \citep{YamPerHoMeoNiiGra}, and is possibly also
responsible for the Soret effect~\citep{DuhBra}. We will, however,
abstain from focusing on these interesting phenomena here, since the
goal of the current work is to address the creation of turbulence
instead.

\section{Numerical simulations of the inertial flow in a straight pipe}
\label{sec:numerical}

Here, we use the inertial flow equations with the hard sphere mean
field potential in \eqref{eq:inertial_flow_HS} to simulate the flow of
air in a straight pipe, with the parameters similar to those in the
experiment by \citet{BucVel}. We use the same computational software
as in our previous works \citep{Abr22,Abr23} -- namely, OpenFOAM
\citep{WelTabJasFur}. Noting that the inertial flow equations in
\eqref{eq:inertial_flow_HS} comprise a system of nonlinear
conservation laws, we simulate them with the help of the appropriately
modified \texttt{rhoCentralFoam} solver \citep{GreWelGasRee}, which
uses the central scheme of \citet{KurTad} for the numerical finite
volume discretization, with the flux limiter due to
\citet{vanLee}. The time-stepping of the method is adaptive, based on
the 20\% of the maximal Courant number.

\subsection{Details of the computational implementation}
\label{sec:BCIC}

\begin{figure}
\includegraphics[width=\textwidth]{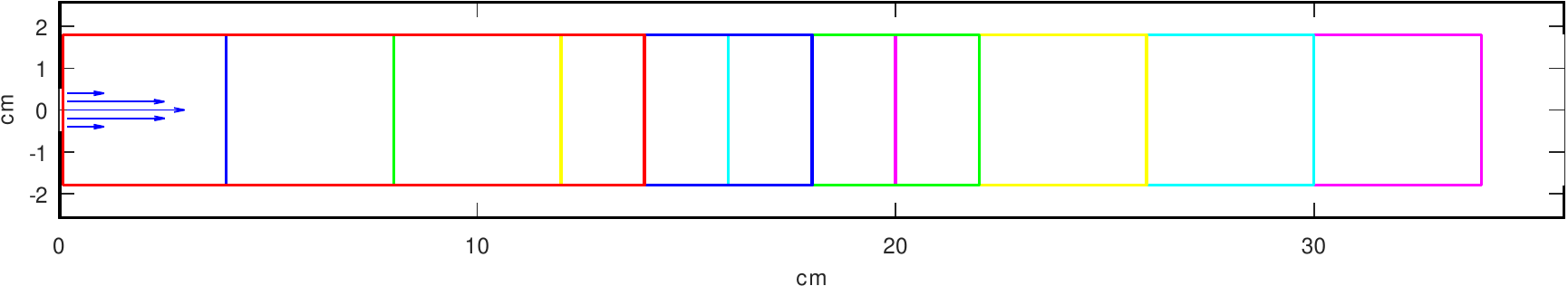}
\caption{The longitudinal section ($z=0$) of the simulation
  domain. The domain dimensions are $36\times 5.2\times 5.2$ cm. The
  inlet is on the left, and the outlet is on the right. The pipe walls
  are shown via thick black lines, so that both the inlet and outlet
  are visible. The boundaries of the Fourier spectrum measurement
  regions are shown in different colors. Each region is a box of 14 cm
  in length, and $3.6\times 3.6$ cm in cross-section. The longitudinal
  offsets of regions are 0 (red), 4 (blue), 8 (green), 12 (yellow), 16
  (cyan) and 20 (magenta) cm from the inlet.}
\label{fig:domain}
\end{figure}

We simulate the inertial flow equations with the hard sphere mean
field potential \eqref{eq:inertial_flow_HS} in a straight pipe of a
square cross-section, with dimensions of $36\times 5.2\times 5.2$
cm. In the corresponding Cartesian reference frame, the $x$-coordinate
varies between 0 and 36 cm, while both the $y$- and $z$-coordinates
vary between $-2.6$ and $2.6$ cm. The longitudinal section of the
domain (that is, what lies in the $xy$-plane with $z=0$) is shown in
Figure \ref{fig:domain}.  The spatial discretization step is $0.8$ mm
in all three dimensions, totaling $450\times 65\times 65=1,901,250$
finite volume cells.

The air enters the pipe through the round inlet of 1 cm in diameter,
located in the middle of the square wall which lies in the $yz$-plane
(that is $x=0$ cm). The air exits the pipe through a square $3.6\times
3.6$ cm outlet, located in the middle of the opposite square wall at
$x=36$ cm (so that, within the outlet, both the $y$- and $z$-
coordinates vary between $-1.8$ and $1.8$ cm). Both the inlet and
outlet are visible in Figure \ref{fig:domain} as gaps on the left- and
right-hand sides of the plot. The length of the domain, together with
the size and shape of the inlet, are consistent with those in the
experiment of \citet{BucVel}.

In what follows, we compute the Fourier spectra of the streamwise
kinetic energy and temperature in the six regions, distributed
uniformly along the pipe. Each region is a box of 14 cm in length and
$3.6\times 3.6$ cm in cross-section. The regions are situated on the
longitudinal axis of the pipe with varying distances from the inlet
wall: 0, 4, 8, 12, 16 and 20 cm. These regions are also displayed in
Figure \ref{fig:domain} in different colors.

The constant parameters $p_0$ and $\rho_\HS$ in
\eqref{eq:inertial_flow_HS} are set, respectively, to 10$^5$ Pa
(normal pressure), and 1850 kg/m$^3$, based on the molar mass and
viscosity of air (for details, see \citet{Abr23}). The following
boundary conditions are used throughout the computation:
\begin{itemize}
\item {\bf Inlet and walls:} For the density, we set the Neumann
  boundary condition with zero normal derivative. For the velocity, we
  set the Dirichlet boundary condition, with zero value at the walls,
  and a radially symmetric parabolic profile at the inlet, with the
  maximum value of 30 m/s at the center (as in the experiment of
  \citet{BucVel}), directed along the $x$-axis of the pipe.
\item {\bf Outlet:} For the density, we set the Dirichlet boundary
  condition with the value of $1.209$ kg/m$^3$, which corresponds to
  the density of air at normal conditions (that is, the pressure
  $10^5$ Pa and temperature 288.15 K, or 15$^\circ$ C). For the
  velocity, we set the Neumann boundary condition with zero normal
  derivative.
\end{itemize}
At the start of the numerical simulation, the air inside the domain is
at rest (zero velocity, $1.209$ kg/m$^3$ density). This initial
condition simulates a realistic laboratory scenario, where a jet enters
the otherwise resting air.  This is markedly different from what we
used previously in~\citep{Abr22,Abr23}, where the initial condition
was a laminar jet inside the domain.

The reason why the outlet in our domain does not comprise the whole
square wall, but rather forms a ``window'' in it, is the following.
Given that the mathematical properties of \eqref{eq:inertial_flow_HS}
have not been studied in detail as of yet, the consistency of its
boundary conditions is currently an open question. However,
empirically, we found that, when the adjacent domain patches with
different types of boundary conditions were transversal, a numerical
instability developed in the vicinity of the conjoining
edge. Conversely, we also found that, when such adjacent patches were
located in the same plane, such numerical instability did not
manifest. Thus, we implemented the window-type outlet at the end of
the pipe to avoid the aforementioned numerical instability.

\subsection{Results of the numerical simulation}

\begin{figure}[t]
\includegraphics[width=\textwidth]{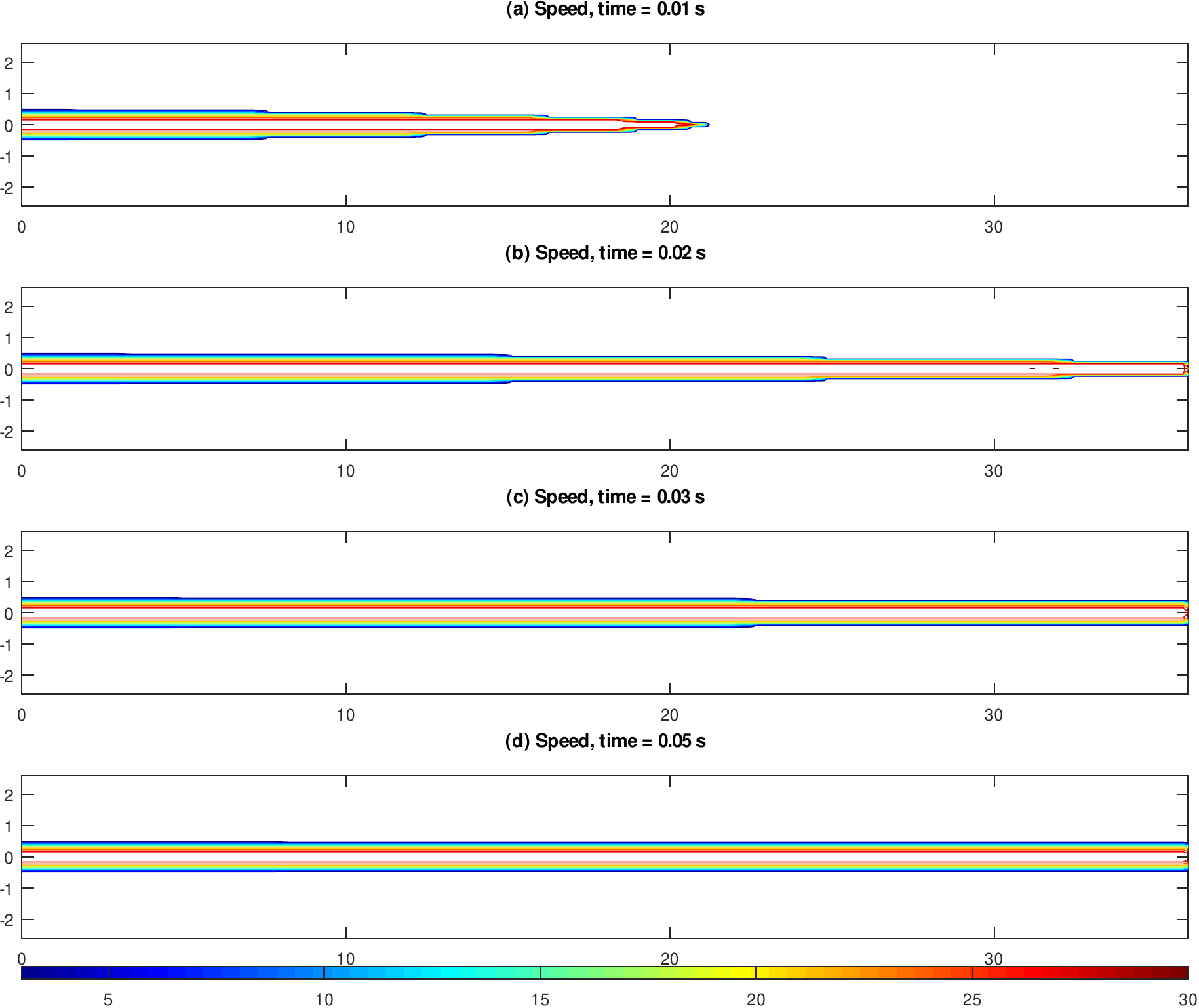}
\caption{Speed of the flow (m/s) without the mean field potential,
  captured in the $xy$-plane ($z=0$) of the pipe at ({\bf a}) 0.01,
  ({\bf b}) 0.02, ({\bf c}) 0.03 and ({\bf d}) 0.05 s of elapsed model
  time.}
\label{fig:speed_no_potential}
\end{figure}

\begin{figure}[t]
\includegraphics[width=\textwidth]{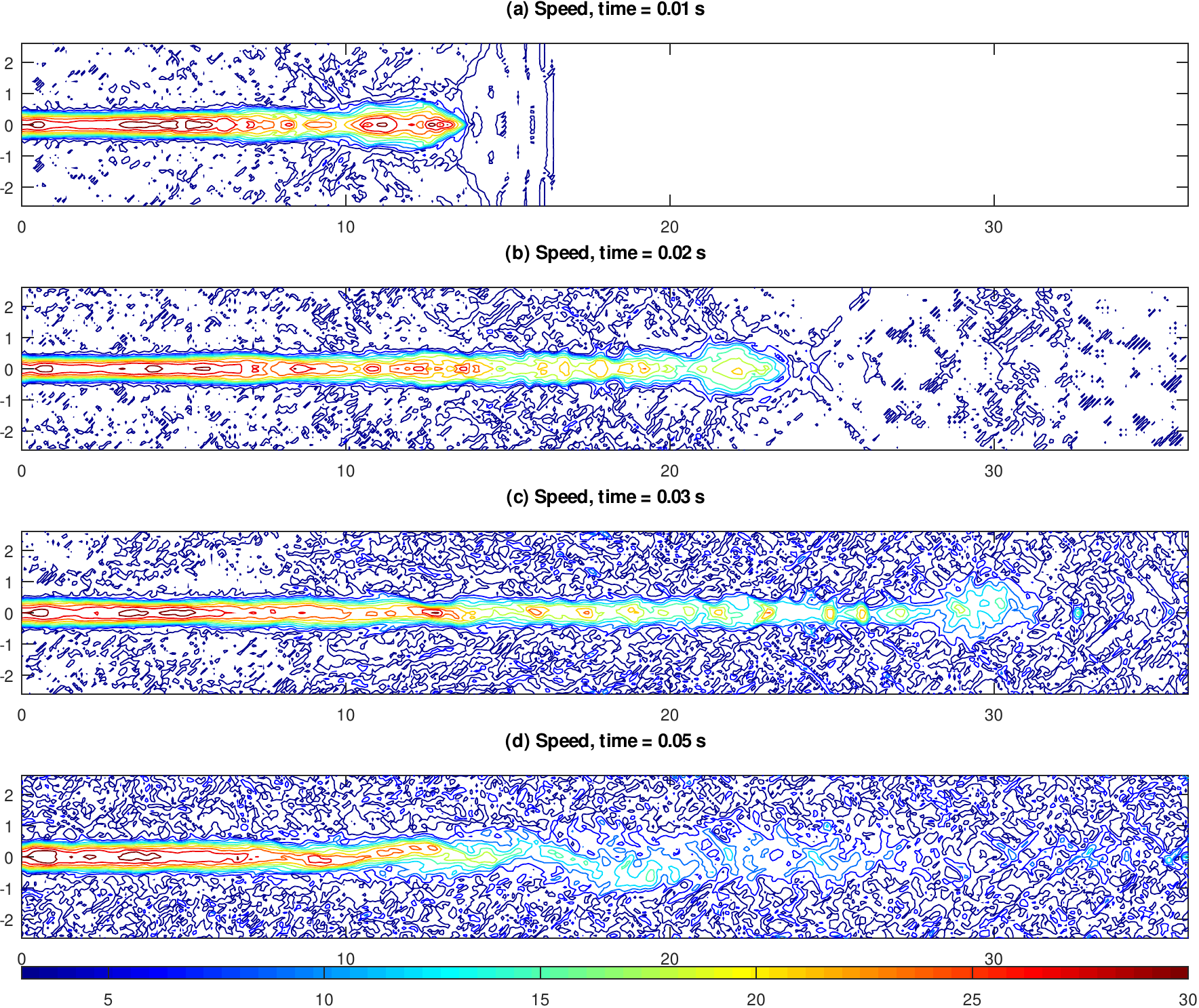}
\caption{Speed of the flow (m/s), captured in the $xy$-plane ($z=0$)
  of the pipe at ({\bf a}) 0.01, ({\bf b}) 0.02, ({\bf c}) 0.03 and
  ({\bf d}) 0.05 s of elapsed model time.}
\label{fig:speed}
\end{figure}

As a benchmark, we first simulated the inertial flow equations in
\eqref{eq:inertial_flow_HS} without the mean field potential. In such
a case, the velocity becomes decoupled from the density, and is
governed by the stand-alone equation
\begin{equation}
\parderiv{\BV u}t+(\BV u\cdot\nabla)\BV u=\BV 0.
\end{equation}
The above equation does not, however, have the form of a conservation
law, and thus the numerical simulation is still completed using
\eqref{eq:inertial_flow_HS} with the initial and boundary conditions
set as described above in Section \ref{sec:BCIC}, except that the mean
field potential forcing is set to zero. In the same fashion as we did
in \citep{Abr22}, here we show the snapshots of the speed of the flow,
captured in the $xy$-plane of the pipe (that is, for $z=0$), in the
form of contour plots. The snapshots are taken at the times $t=0.01$,
$0.02$, $0.03$ and $0.05$ seconds of the elapsed time, and are shown
in Figure~\ref{fig:speed_no_potential}. As we can see, in the absence
of the potential forcing, the jet ``pierces'' the resting air and
exits through the outlet, remaining laminar in the process. Once the
jet fully traverses the length of the pipe, the numerical solution
reaches a steady state. The air outside of the jet (that is, at the
distance more than 0.5 cm from the axis of the pipe) remains at rest
at all times.

For comparison, in Figure \ref{fig:speed}, we show the speed of the
flow in the same form as in Figure~\ref{fig:speed_no_potential},
except that the potential forcing is set as described in Section
\ref{sec:BCIC}, with $p_0=10^5$ Pa and $\rho_\HS=1850$ kg/m$^3$. As
we can see, there is a remarkable difference between the plots. Once
the potential forcing is present, the jet creates fluctuations around
itself, disintegrating in the process, just as typically observed in
nature and experiments \citep{Rey83,BucVel}, as well as in numerical
simulations of our recent works \citep{Abr22,Abr23}. The length of the
pipe is such that the jet completely falls apart by the time the flow
reaches the outlet -- in fact, at time $t=0.03$ s, the jet extends
somewhat farther than in the subsequent snapshot, taken at time
$t=0.05$ s. By the time $t=0.05$ s, the structure of the flow is as
follows. In the first third of the pipe, the jet is largely intact,
with small fluctuations around it.  In the second third of the pipe,
the jet disintegrates, transitioning into a chaotic flow. In the last
third of the pipe, the flow is completely chaotic, with weak remnants
of the jet, which can be observed in the middle of the pipe. After the
time $t=0.05$ s, the overall structure of the flow remained the same
as the computation proceeded, and thus we concluded that the
statistical steady state has been reached at $t=0.05$.  However,
unlike the simulation without the potential shown in Figure
\ref{fig:speed_no_potential}, here the flow never settles on a steady
state, and remains in chaotic motion at all times.

The remarkable difference between the flows in Figures
\ref{fig:speed_no_potential} and \ref{fig:speed} confirms that, at
normal conditions, the effect of an intermolecular potential,
expressed via the mean field forcing, is non-negligible and
quantifiable. Furthermore, the presence of the mean field potential
unambiguously creates turbulent motions in the numerical simulation of
an inertial flow, and these motions are similar to those observed in
nature and experiments.

In addition, observe that the geometry of the domain, as well as the
initial and boundary conditions, are symmetric with respect to the
central axis of the pipe. This means that, technically, the flow must
also remain symmetric relative to the central axis at all
times. However, it remains symmetric only in the absence of the
intermolecular potential (see Figure
\ref{fig:speed_no_potential}). When the intermolecular potential is
present, we observe that the flow symmetry breaks as early as at
$t=0.02$ seconds (plot ({\bf b}) in Figure \ref{fig:speed}, small
fluctuations near the outlet). This breaking of the symmetry is a
reliable indication of a strongly chaotic dynamics -- what we observe
is round-off errors of the floating point arithmetic, which introduce
a very weak asymmetry into the numerical solution, grow exponentially
rapidly in time due to large positive Lyapunov exponents. As an
example, a similar phenomenon manifested in the work of
\citet{MajTim}, where the initial condition from a low-dimensional
stable manifold of the truncated Burgers--Hopf system evolved into a
fully chaotic and mixing solution due to the exponential growth of
machine round-off errors (for a more detailed explanation, see
\citep{AbrPhDthesis}, Section 2.2.4).

\begin{figure}[t]
\includegraphics[width=\textwidth]{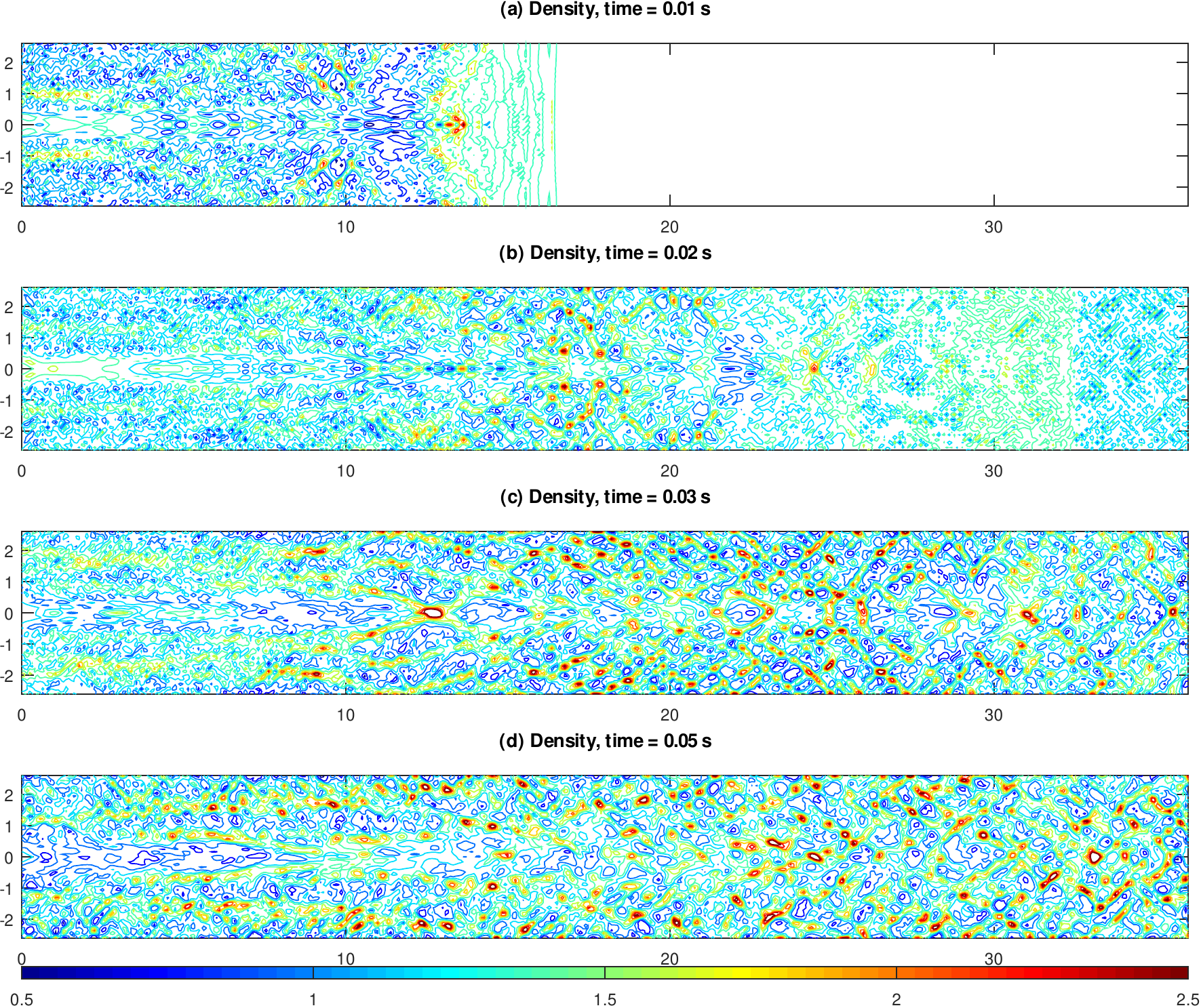}
\caption{Density of the flow (kg/m$^3$), captured in the $xy$-plane
  ($z=0$) of the pipe at ({\bf a}) 0.01, ({\bf b}) 0.02, ({\bf c})
  0.03 and ({\bf d}) 0.05 s of elapsed model time.}
\label{fig:density}
\end{figure}

\begin{figure}[t]
\includegraphics[width=\textwidth]{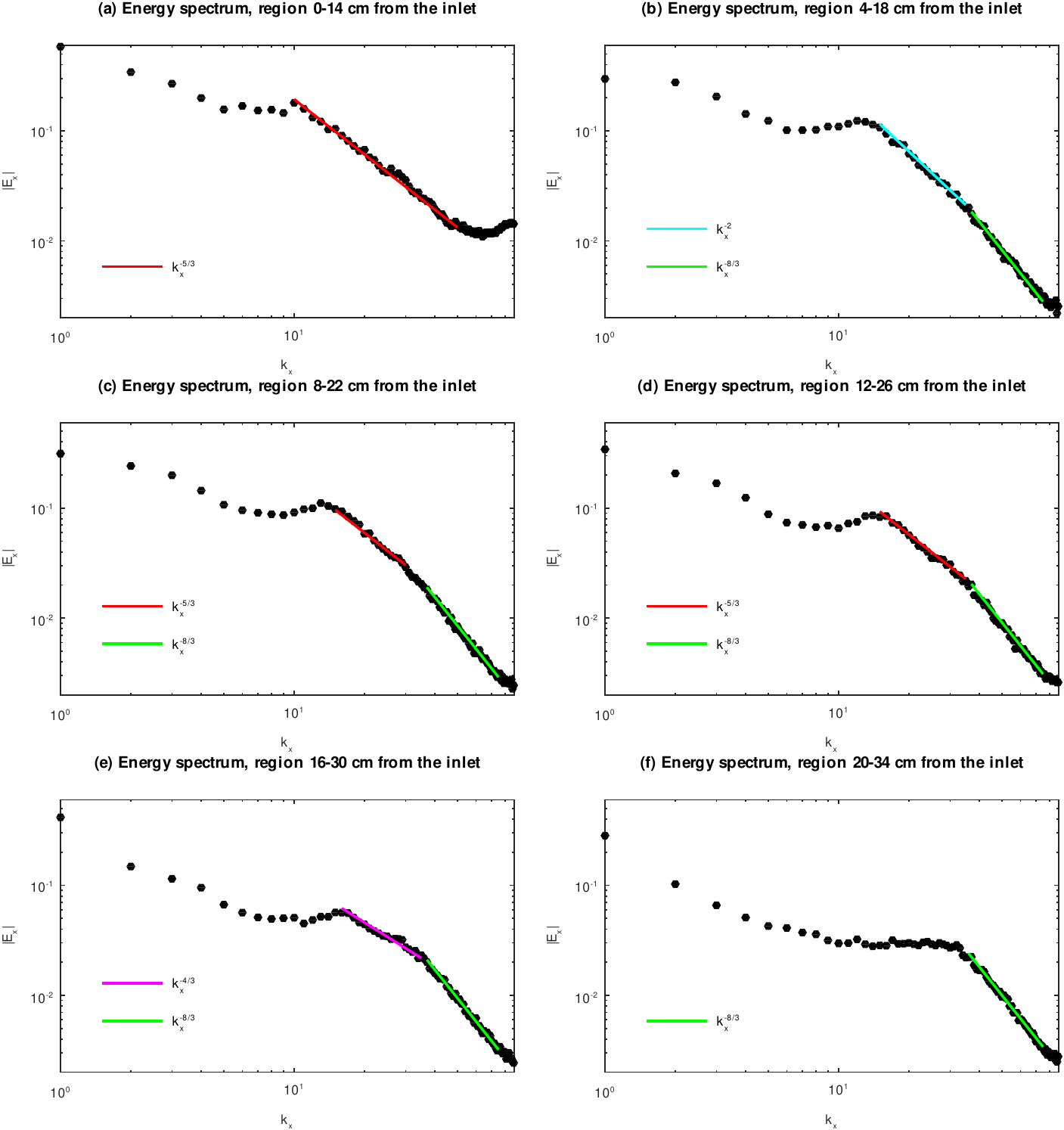}
\caption{The Fourier spectra of the kinetic energy, computed within
  different spatial windows, situated at ({\bf a}) 0--14, ({\bf b})
  4--18, ({\bf c}) 8--22, ({\bf d}) 12--26, ({\bf e}) 16--30 and ({\bf
    f}) 20--34 cm, counting from the inlet. The slope lines
  $k^{-5/3}$, $k^{-2}$, $k^{-4/3}$ and $k^{-8/3}$ are given for the
  reference.}
\label{fig:energy_spectrum}
\end{figure}

\subsubsection{The density of the flow}

In addition to the snapshots of the flow speed, in
Figure~\ref{fig:density} we show the snapshots of the density of the
flow, captured in the same way and at the same times, and also
displayed as contour plots. Just as in our recent works
\citep{Abr22,Abr23}, observe that the fluctuations of density from its
background value of $1.209$ kg/m$^3$ are rather unrealistic -- as much
as by a factor of two both ways. We currently presume that this
happens due to that fact that our model is ``too idealized''; namely,
the pressure is presumed to be strictly constant irrespectively of how
the other variables behave. In a real gas, obviously, such condition
may not strictly hold -- if a large enough density gradient develops
locally in a flow, it would also inevitably create a pressure
fluctuation. Our simplified model, on the other hand, adheres strictly
to Charles' law for a constant pressure process, and thus has certain
bounds of practical applicability.  Generally, it appears to be a
challenging problem to create a universal gas transport model, which
would be accurate in all thermodynamic regimes of the flow, from
Charles' law at low Mach numbers, to acoustic waves and adiabatic
shock transitions at high Mach numbers.

\subsubsection{The Fourier spectra of the kinetic energy}

Here, we show the time averages of the Fourier spectra of the
streamwise kinetic energy of the flow (that is, the energy of the
$x$-component of the velocity), computed in the six regions which were
described in Section \ref{sec:BCIC} and shown in Figure
\ref{fig:domain}.  In each region, the kinetic energy spectrum was
computed as follows: first, the kinetic energy of the $x$-component of
the velocity, $E_x=u_x^2/2$, was averaged over the cross-section of
the region, thus becoming the function of the $x$-coordinate
only. Then, the linear trend was subtracted from the result in the
same manner as was done by \citet{NasGag} and also in our recent works
\citep{Abr22,Abr23}, to ensure that there was no sharp discontinuity
between the energy values at the western and eastern boundaries of the
region. Finally, the one-dimensional discrete Fourier transformation
was applied to the result. The subsequent time-averaging of the
modulus of the Fourier transform was computed in the time interval
$0.1\leq t\leq 0.2$ seconds of the elapsed model time.

\begin{figure}[t]
\includegraphics[width=\textwidth]{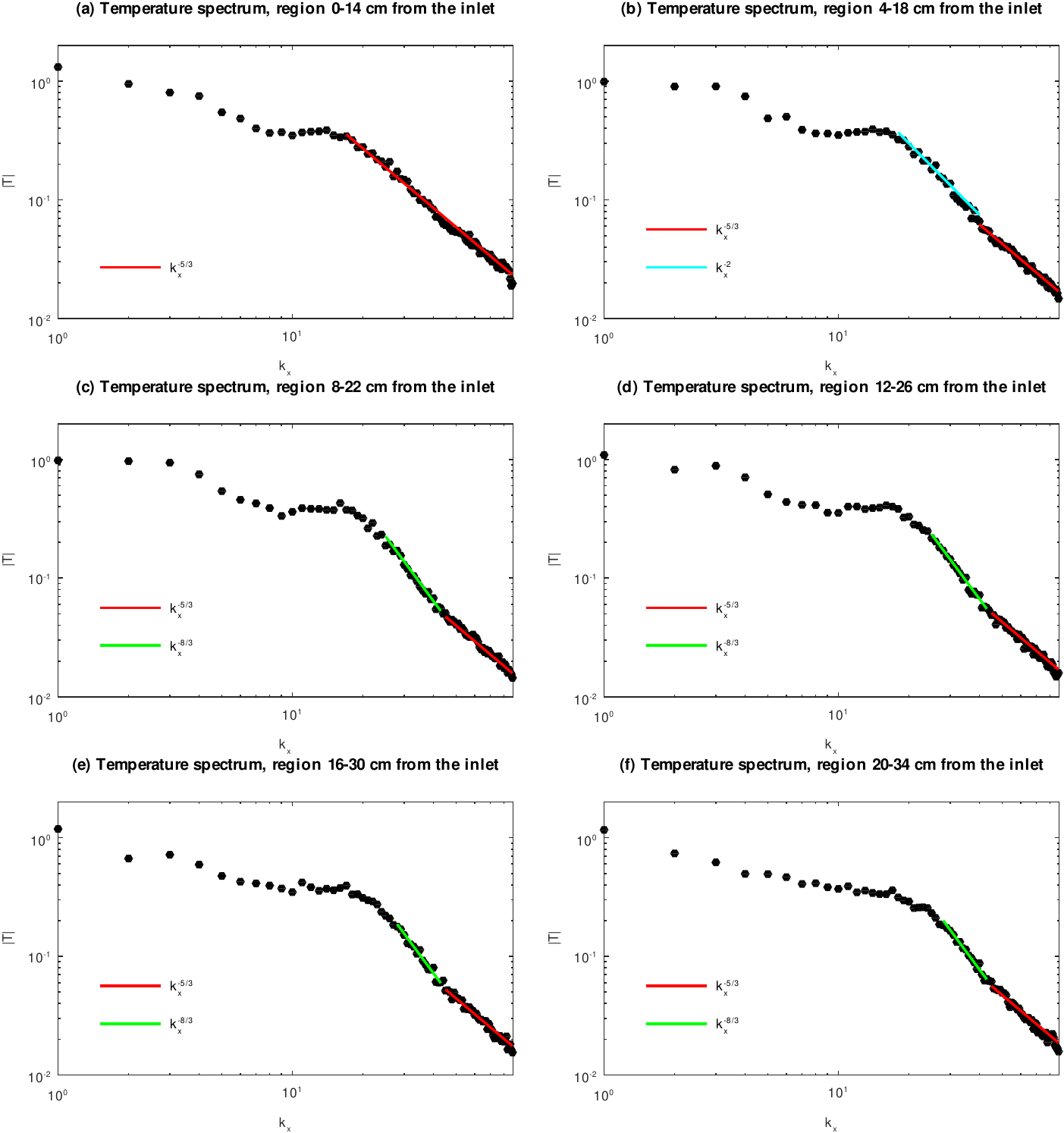}
\caption{The Fourier spectra of the temperature, computed within
  different spatial windows, situated at ({\bf a}) 0--14, ({\bf b})
  4--18, ({\bf c}) 8--22, ({\bf d}) 12--26, ({\bf e}) 16--30 and ({\bf
    f}) 20--34 cm, counting from the inlet. The slope lines
  $k^{-5/3}$, $k^{-2}$ and $k^{-8/3}$ are given for the reference.}
\label{fig:temperature_spectrum}
\end{figure}

The time averages of the kinetic energy spectra, computed as described
above, are shown in Figure \ref{fig:energy_spectrum} for all six
regions, in the ascending order of their distance from the inlet. In
the first region, which begins directly at the inlet (plot ({\bf a})
in Figure \ref{fig:energy_spectrum}), the kinetic energy spectrum has
the decay of $k_x^{-5/3}$ (the famous Kolmogorov spectrum) on the
moderate and small scales, and looks very similar to what was observed
by \citet{BucVel} in their experiment, and simulated in our recent
work \citep{Abr22} for the flow of argon. However, as the measurement
region shifts away from the inlet, a qualitative change in the kinetic
energy spectra is observed -- the decay rates become different between
the moderate and small scales.  At the small scales, starting
approximately with the wavenumber 35 and up, the decay rate of the
kinetic energy corresponds to $k_x^{-8/3}$ in all remaining
measurement regions, and looks similar to what we observed in our
recent work \citep{Abr23} for a large scale two-dimensional
flow. However, at the moderate scales (between the wavenumbers 15 and
35, approximately) the power of the decay rate varies between
different regions:
\begin{itemize}
\item In the region beginning at 4 cm from the outlet (plot ({\bf
  b}) in Figure \ref{fig:energy_spectrum}), the rate of decay
  corresponds to $k_x^{-2}$;
\item In the regions beginning at 8 and 12 cm from the outlet (plots
  ({\bf c}) and ({\bf d}) in Figure~\ref{fig:energy_spectrum}), the
  rate of decay corresponds to $k_x^{-5/3}$;
\item In the region beginning at 16 cm from the outlet (plot ({\bf
  e}) in Figure \ref{fig:energy_spectrum}), the rate of decay becomes
  $k_x^{-1}$;
\item In the region beginning at 20 cm from the outlet (plot ({\bf f})
  in Figure \ref{fig:energy_spectrum}), there is no decay at the
  moderate scales -- the spectrum is flat.
\end{itemize}
Such two-tiered kinetic energy spectra, with steeper decay at small
scales, were observed in the Jovian atmosphere by \textit{Cassini}
\citep{ChoSho} and \textit{Juno} \citep{MorMigAlt} missions.

\subsubsection{The Fourier spectra of the temperature}

In addition to the kinetic energy spectra of the large scale
meridional and zonal winds, \citet{NasGag} reported the Fourier
spectra of the temperature, and found that the latter also had power
scaling. Therefore, in the present work, we also report the
time-averaged Fourier spectra of the temperature, which are computed
in the precisely same manner as those of the kinetic energy, presented
above. The temperature $T$ (in $^\circ$K) is computed via the formula
\begin{equation}
T=\frac MR\frac{p_0}\rho,
\end{equation}
where $R=8.31446$ kg m$^2$/s$^2$ mol K is the universal gas constant,
and $M=2.897$ kg/mol is the molar mass of air.

We show the temperature spectra in Figure
\ref{fig:temperature_spectrum}, computed in the same measurement
regions as shown in Figure \ref{fig:domain}. As we can see, in the
first measurement region, which extends from 0 to 14 cm, counting from
the inlet, the rate of decay of the temperature spectrum largely
corresponds to $k_x^{-5/3}$ (plot ({\bf a}) in Figure
\ref{fig:temperature_spectrum}). However, a ``two-tiered'' decay is
observed in the rest of the measurement regions, where, contrary to
what was observed above for the kinetic energy, the steeper decay of
$k_x^{-2}$ and $k_x^{-8/3}$ is observed at the moderate scales
(between the wavenumbers 25 and 45, approximately). At small scales,
the temperature spectrum decay corresponds to $k_x^{-5/3}$ for all
measurement regions. We note that this power structure of the
temperature spectrum (with slower decay rate at smaller scales) is
similar to what was observed by \citet{NasGag}.

\section{Summary}
\label{sec:summary}

In the current work, we extend our previous results in \citep{Abr22}
and \cite{Abr23} onto polyatomic gas flows. We develop a tractable
kinetic model of a general polyatomic gas, which describes the
translational interaction via a deterministic intermolecular
potential, and rotational interactions via stochastic collisions. In
the suitable hydrodynamic limit for the distribution function of a
single particle, we arrive at what we refer to as the
Boltzmann--Vlasov equation, because our kinetic equation possesses
both the deterministic potential and the stochastic collision
integral. For the transport equations of velocity moments, we
introduce a novel heat flux closure by prescribing the appropriate
specific heat capacity of the process. For the specific heat capacity
taken at a constant pressure, we obtain the inertial flow equations of
the same form as in \citep{Abr22}, albeit with a slightly more
accurate estimate of the mean field potential, which now includes the
cavity distribution function.

For the numerical simulation, we choose a scenario which roughly
corresponds to the experiment of \citet{BucVel}. The air at normal
conditions enters a straight pipe through the inlet of 1 cm in
diameter, with the parabolic velocity profile at the maximum speed of
30 m/s. We show that, in the absence of an interaction potential, the
resulting laminar jet ``pierces'' the resting air and exits through
the outlet; yet, in the presence of the interaction potential, the jet
quickly breaks up into turbulent motions similarly to the results in
our recent works \citep{Abr22,Abr23}. We record the kinetic energy
spectra within measurement regions at different distances from the
inlet, and find a variety of different power spectra (in particular, a
double-slope decay similar to that observed on Jupiter
\citep{ChoSho,MorMigAlt}).  We also measure the temperature spectra of
the flow in the same fashion, and observe that the structure of the
power decay is similar to that observed in the Earth atmosphere
\citep{NasGag}.

\ack This work was supported by the Simons Foundation grant \#636144.

\appendix

\section{The entropy inequality for the forward Kolmogorov equation}
\label{sec:entropy_inequality}

To obtain the entropy inequalities in \eqref{eq:entropy} for the
forward Kolmogorov equation \eqref{eq:kolmogorov}, let us consider the
quantity
\begin{equation}
Q[\Psi_1(F),\Psi_2(F_0)]=\int \Psi_1(F)\Psi_2(F_0)\dif\BV X\dif\BV V,
\end{equation}
where $F_0$ is a steady state of \eqref{eq:kolmogorov}, and
$\Psi_1:\RR\to\RR$ and $\Psi_2:\RR\to\RR$ are two differentiable
functions. The time derivative of $Q$ is given via
\begin{multline}
\parderiv{}t Q[\Psi_1(F),\Psi_2(F_0)]=\int D\Psi_1(F)\parderiv Ft
\Psi_2(F_0)\dif\BV X\dif\BV V=\int D\Psi_1(F)\Psi_2(F_0)\\\bigg[
  \parderiv\Phi{\BV X}\cdot\parderiv F{\BV V}-\BV V\cdot\parderiv
  F{\BV X}+\sum_{i=1}^{K-1}\sum_{j=i+1}^K \lambda_{ij} \big(F(\collmap_{i
    j}^{-1}\BV V)-F(\BV V)\big)\bigg]\dif\BV X \dif\BV V,
\end{multline}
where the notation $D\Psi_1$ denotes the derivative of $\Psi_1$ with
respect to its argument. Observe that the part of the latter integral
which does not involve collisions is zero; indeed, the integration by
parts yields,
\begin{multline}
\int D\Psi_1(F)\Psi_2(F_0)\left(\parderiv\Phi{\BV X}\cdot\parderiv F{
  \BV V}-\BV V\cdot\parderiv F{\BV X}\right)\dif\BV X\dif\BV V\\=\int
\Psi_1(F)D\Psi_2(F_0)\left(\BV V\cdot\parderiv{F_0}{\BV X}- \parderiv
\Phi{\BV X}\cdot\parderiv{F_0}{\BV V}\right)\dif\BV X\dif\BV V=0.
\end{multline}
For the collision term, we use the fact that both $\lambda_{ij}$ and
$F_0$ are invariant under $\collmap_{ij}$, and obtain
\begin{equation}
\parderiv{}t Q[\Psi_1(F),\Psi_2(F_0)]=\sum_{i=1}^{K-1}\sum_{j=i+1}^K
\int \lambda_{ij} F\big[D\Psi_1(F(\collmap_{ij}(\BV V)))-D\Psi_1(F(\BV
  V)) \big] \Psi_2(F_0)\dif\BV X\dif\BV V.
\end{equation}
Clearly, for some special cases of $\Psi_1$ and $\Psi_2$, the
expression above can be treated further. The most obvious
simplification occurs when $\Psi_1(F)=F$, in which case the expression
above is zero irrespective of $\Psi_2$:
\begin{equation}
\label{eq:Qzero}
\parderiv{}t Q[F,\Psi_2(F_0)]=0.
\end{equation}
Next, let us consider the special case with $\Psi_1(F)=-F\ln F$,
$\Psi_2(F_0)=1$, such that $Q$ is the Shannon entropy. In this case,
we prove the first statement in \eqref{eq:entropy}:
\begin{multline}
\label{eq:shannon}
-\parderiv{}t\int F\ln F\dif\BV X\dif\BV V=\sum_{i=1}^{K-1}\sum_{j=i+1}
^K\int\lambda_{ij}F(\BV V)\ln\bigg(\frac{F(\BV V)}{F(\collmap_{ij}(\BV V)
  )}\bigg)\dif\BV X\dif\BV V\\\geq\sum_{i=1}^{K-1}\sum_{j=i+1}^K\int
\lambda_{ij}F(\collmap_{ij}(\BV V))\bigg(\frac{F(\BV V)}{F(\collmap_{ij}
  (\BV V))}-1\bigg)\dif\BV X\dif\BV V=0,
\end{multline}
where we used the inequality $x\ln x\geq x-1$. Setting
$\Psi_2(F_0)=\ln F_0$ in \eqref{eq:Qzero}, and subtracting
\eqref{eq:shannon}, yields the second statement in \eqref{eq:entropy}.

\section{The hydrodynamic limit for the Boltzmann--Vlasov equation}
\label{app:hydrodynamic_limit}

Here we compute the integral in the right-hand side of
\eqref{eq:transport}, with the rescalings in \eqref{eq:rescaling}.

\subsection{Potential forcing}

Upon replacing the dummy variable of integration $\BV x_2=\BV
x-\sigma\BV r$, the integral for the potential forcing in the
right-hand side of \eqref{eq:transport} becomes
\begin{multline}
\label{eq:forc_integral}
\frac 1m\int e^{-\frac{\phi(\|\BV x-\BV x_2\|/\sigma)}{\theta((\BV
    x+\BV x_2)/2)}} Y_K(\theta((\BV x+\BV x_2)/2),\|\BV x-\BV
x_2\|/\sigma) \parderiv{}{\BV x}\phi(\|\BV x-\BV x_2\|/\sigma)\rho(\BV
x_2)\dif\BV x_2 \\=\frac{\sigma^2}m\int e^{-\frac{\phi(\|\BV
    r\|)}{\theta(\BV x- \sigma\BV r/2)}} Y_K(\theta(\BV x-\sigma\BV
r/2),\|\BV r\|)\parderiv{\phi(\|\BV r\|)}{\BV r}\rho(\BV x-\sigma\BV
r)\dif\BV r.
\end{multline}
Next, we observe that
\begin{multline}
e^{-\frac{\phi(\|\BV r\|)}{\theta(\BV x-\sigma\BV r/2)}} \parderiv{
  \phi(\|\BV r\|)}{\BV r}=e^{-\frac{\phi(\|\BV r\|)}{\theta(\BV x-
    \sigma\BV r/2)}}\theta(\BV x-\sigma\BV r/2)\left(\frac\sigma
2\parderiv{}{\BV x}+\parderiv{}{\BV r}\right)\left(\frac{\phi(\|\BV
  r\|)}{\theta(\BV x-\sigma\BV r/2)}\right)\\=\theta(\BV x-\sigma\BV
r/2)\left(\frac\sigma 2\parderiv{}{\BV x}+\parderiv{}{\BV
  r}\right)\left(1-e^{-\frac{\phi(\|\BV r\|)}{\theta(\BV x-\sigma\BV
    r/2)}}\right).
\end{multline}
We now can now manipulate the integral as follows:
\begin{multline}
\frac{\sigma^2}m\int e^{-\frac{\phi(\|\BV r\|)}{\theta(\BV x-
    \sigma\BV r/2)}} Y_K(\theta(\BV x-\sigma\BV r/2),\|\BV
r\|)\parderiv{\phi(\|\BV r\|)}{\BV r}\rho(\BV x-\sigma\BV r)\dif\BV
r\\=\frac{\sigma^3}{2m}\int Y_K(\theta(\BV x-\sigma\BV r/2),\|\BV
r\|)\theta(\BV x-\sigma\BV r/2) \rho(\BV x-\sigma\BV r)\parderiv{}{\BV
  x}\left(1-e^{-\frac{\phi(\|\BV r \|)}{\theta(\BV x-\sigma\BV
    r/2)}}\right)\dif\BV r\\+\frac{\sigma^2}m \int Y_K(\theta(\BV
x-\sigma\BV r/2),\|\BV r\|)\theta(\BV x-\sigma\BV r/ 2)\rho(\BV
x-\sigma\BV r)\parderiv{}{\BV r}\left(1-e^{-\frac{ \phi(\|\BV
    r\|)}{\theta(\BV x-\sigma\BV r/2)}}\right)\dif\BV r.
\end{multline}
The first integral above can be expressed via
\begin{multline}
\frac{\sigma^3}{2m}\int Y_K(\theta(\BV x-\sigma\BV r/2),\|\BV r\|)
\theta(\BV x-\sigma\BV r/2)\rho(\BV x-\sigma\BV r)\parderiv{}{\BV x}
\left(1-e^{-\frac{\phi(\|\BV r\|)}{\theta(\BV x-\sigma\BV r/2)}}
\right)\dif\BV r\\=\frac{\sigma^3}{2m}\parderiv{}{\BV x}\int
Y_K(\theta(\BV x-\sigma\BV r/2),\|\BV r\|)\theta(\BV x-\sigma\BV
r/2)\rho(\BV x- \sigma\BV r)\left(1-e^{-\frac{\phi(\|\BV
    r\|)}{\theta(\BV x-\sigma\BV r/2)}}\right)\dif\BV
r\\-\frac{\sigma^3}{2m}\int\left(1-e^{-\frac{ \phi(\|\BV
    r\|)}{\theta(\BV x-\sigma\BV r/2)}}\right)\parderiv{}{ \BV
  x}\big[Y_K(\theta(\BV x-\sigma\BV r/2),\|\BV r\|)\theta(\BV
  x-\sigma\BV r/2)\rho(\BV x-\sigma\BV r)\big]\dif\BV r\\=\frac{
  \sigma^3}{2m}\parderiv{}{\BV x}\int Y_K(\theta(\BV x-\sigma\BV r/2),
\|\BV r\|)\theta(\BV x-\sigma\BV r/2)\rho(\BV x-\sigma\BV r)\left( 1-
e^{-\frac{\phi(\|\BV r\|)}{\theta(\BV x-\sigma\BV r/2)}}\right)\dif\BV
r\\+\frac{\sigma^2}m\int\left(1-e^{-\frac{\phi(\|\BV r\|)}{\theta(\BV
    x-\sigma \BV r/2)}}\right)\left(-\frac\sigma
2\right)\parderiv{}{\BV x} \big[Y_K(\theta(\BV x-\sigma\BV r/2),\|\BV
  r\|)\theta(\BV x-\sigma\BV r/2)\rho(\BV x-\sigma\BV r)\big]\dif\BV
r,
\end{multline}
where we observe that
\begin{multline}
-\frac\sigma 2 \parderiv{}{\BV x}\big[Y_K(\theta(\BV x-\sigma\BV
  r/2),\|\BV r\|)\theta(\BV x-\sigma\BV r/2)\rho(\BV x-\sigma\BV
  r)\big]\\=\parderiv{}{\BV r}\big[Y_K(\theta(\BV x-\sigma\BV
  r/2),\|\BV r\|)\theta(\BV x-\sigma\BV r/2)\rho(\BV x-\sigma\BV
  r)\big]\\-\theta(\BV x-\sigma\BV r/2)\rho(\BV x-\sigma\BV
r)\parderiv{}r Y_K(\theta(\BV x-\sigma\BV r/2),\|\BV r\|)\frac{\BV
  r}{\|\BV r\|}\\+\frac\sigma 2Y_K(\theta(\BV x-\sigma\BV r/2),\|\BV
r\|)\theta(\BV x-\sigma\BV r/2)\parderiv{}{\BV x}\rho(\BV
x-\sigma\BV r).
\end{multline}
The second integral can be integrated by parts:
\begin{multline}
\frac{\sigma^2}m\int Y_K(\theta(\BV x-\sigma\BV r/2),\|\BV
r\|)\theta(\BV x- \sigma\BV r/2)\rho(\BV x-\sigma\BV r)\parderiv{}{\BV
  r}\left(1- e^{-\frac{\phi(\|\BV r\|)}{\theta(\BV x-\sigma\BV
    r/2)}}\right)\dif\BV
r\\=-\frac{\sigma^2}m\int\left(1-e^{-\frac{\phi(\|\BV r\|)}{\theta(\BV
    x- \sigma\BV r/2)}}\right)\parderiv{}{\BV r}\big[Y_K(\theta(\BV x-
  \sigma\BV r/2),\|\BV r\|)\theta(\BV x-\sigma\BV r/2)\rho(\BV x-
  \sigma\BV r)\big]\dif\BV r.
\end{multline}
Thus, \eqref{eq:forc_integral} is given via
\begin{multline}
\frac{\sigma^2}m\int e^{-\frac{\phi(\|\BV r\|)}{\theta(\BV x-
    \sigma\BV r/2)}} Y_K(\theta(\BV x-\sigma\BV r/2),\|\BV
r\|)\parderiv{\phi(\|\BV r\|)}{\BV r}\rho(\BV x-\sigma\BV r)\dif\BV
r\\=\frac{\sigma^3}{2m}\parderiv{}{ \BV x}\int Y_K(\theta(\BV
x-\sigma\BV r/2),\|\BV r\|)\theta(\BV x- \sigma\BV r/2)\rho(\BV
x-\sigma\BV r)\left(1-e^{-\frac{\phi(\|\BV r\|) }{\theta(\BV
    x-\sigma\BV r/2)}}\right)\dif\BV r\\+\frac{\sigma^3}{2m} \int
Y_K(\theta(\BV x-\sigma\BV r/2),\|\BV r\|)\theta(\BV x-\sigma\BV r/
2)\left(1-e^{-\frac{\phi(\|\BV r\|)}{\theta(\BV x-\sigma\BV r/2)}}
\right)\parderiv{}{\BV x}\rho(\BV x-\sigma\BV r)\dif\BV r\\-\frac{\sigma^2}m
\int\left(1-e^{-\frac{\phi(\|\BV r\|)}{\theta(\BV x-\sigma\BV r/2)}}
\right)\theta(\BV x-\sigma\BV r/2)\rho(\BV x-\sigma\BV r)\parderiv{}r
Y_K(\theta(\BV x-\sigma\BV r/2),\|\BV r\|)\frac{\BV r}{\|\BV r\|}\dif\BV
r.
\end{multline}
The sum of the first two integrals in the right-hand side above is
\begin{multline}
\frac{\sigma^3}{2m}\parderiv{}{\BV x}\int Y_K(\theta(\BV x-\sigma\BV
r/2), \|\BV r\|)\theta(\BV x-\sigma\BV r/2)\rho(\BV x-\sigma\BV
r)\left( 1-e^{-\frac{\phi(\|\BV r\|)}{\theta(\BV x-\sigma\BV
    r/2)}}\right)\dif \BV r\\+\frac{\sigma^3}{2m}\int Y_K(\theta(\BV
x-\sigma\BV r/2),\|\BV r\|) \theta(\BV x-\sigma\BV
r/2)\left(1-e^{-\frac{\phi(\|\BV r\|)}{ \theta(\BV x-\sigma\BV
    r/2)}}\right)\parderiv{}{\BV x}\rho(\BV x- \sigma\BV r)\dif\BV
r\\=\frac{\sigma^3}{2m}\int\frac 1{\rho(\BV x- \sigma\BV
  r)}\parderiv{}{\BV x}\Big[Y_K(\theta(\BV x-\sigma\BV r/2), \|\BV
  r\|)\theta(\BV x-\sigma\BV r/2)\rho^2(\BV x-\sigma\BV r)
  \left(1-e^{-\frac{\phi(\|\BV r\|)}{\theta(\BV x-\sigma\BV r/2)}}
  \right)\Big]\dif\BV r\\\to 2\pi\frac{\sigma^3}m\frac 1{\rho(\BV
  x)}\parderiv{}{\BV x}\Big[\theta(\BV x)\rho^2(\BV x) \int_0^\infty
  Y(\theta(\BV x),r)\left(1-e^{-\frac{\phi(r)}{\theta(\BV x)}} \right)
  r^2\dif r\Big],
\end{multline}
as $\sigma\to 0$ and $K\to\infty$, with $\sigma^3/m$ being fixed. The
second integral becomes, upon expanding in powers of $\sigma$ and
taking the same limit,
\begin{multline}
-\frac{\sigma^2}m\int\left(1-e^{-\frac{\phi(\|\BV r\|)}{\theta(\BV
    x-\sigma\BV r/2)}}\right)\theta(\BV x-\sigma\BV r/2)\rho(\BV
x-\sigma\BV r)\parderiv{}r Y_K(\theta(\BV x-\sigma\BV r/2),\|\BV
r\|)\frac{\BV r}{\|\BV r\|}\dif\BV
r\\=-\frac{\sigma^2}m\int\left(1-e^{-\frac{\phi(\|\BV r\|)}{\theta(\BV
    x)}}\right)\theta(\BV x)\rho(\BV x)\parderiv{}r Y_K(\theta(\BV
x),\|\BV r\|)\frac{\BV r}{\|\BV r\|}\dif\BV
r\\+\frac{\sigma^3}{2m}\frac 1{\rho(\BV x)}\parderiv{}{\BV
  x}\left[\theta(\BV x)\rho^2(\BV x)\int\left(1-e^{-\frac{\phi(\|\BV
      r\|)}{\theta(\BV x)}} \right) \parderiv{}r Y_K(\theta(\BV x),\|\BV
  r\|)\cdot\frac{\BV r^2}{\|\BV r\|}\dif\BV r\right]+O(\sigma^4/m)
\\\to\frac{\sigma^3}{2m}\frac 1{\rho(\BV x)}\parderiv{}{\BV
  x}\cdot\left[\theta(\BV x)\rho^2(\BV
  x)\int\left(1-e^{-\frac{\phi(r)}{\theta(\BV x)}} \right)
  \parderiv{}r Y(\theta(\BV x),r)r^3\dif r\BV n^2\dif\BV n\right],
\end{multline}
where $\BV n$ is a unit vector, and the integration over $\dif\BV n$
occurs over a unit sphere.  We recognize that
\begin{equation}
\int\BV n^2\dif\BV n=\frac{4\pi}3\BM I,
\end{equation}
and thus
\begin{multline}
\frac{\sigma^3}{2m}\frac 1{\rho(\BV x)}\parderiv{}{\BV x}\cdot\left[
  \theta(\BV x)\rho^2(\BV x)\int\left(1-e^{-\frac{\phi(r)}{\theta(\BV
      x)}} \right) \parderiv{}r Y(\theta(\BV x),r)r^3\dif r\BV
  n^2\dif\BV n\right]\\=2\pi\frac{\sigma^3}m\parderiv{}{\BV x}\left[
  \theta(\BV x)\rho^2(\BV
  x)\int_0^\infty\left(1-e^{-\frac{\phi(r)}{\theta(\BV x)}}
  \right)\frac{r^3}3 \parderiv{}r Y(\theta(\BV x),r)\dif r\right].
\end{multline}
The sum is, therefore,
\begin{equation}
\frac{2\pi}3\frac{\sigma^3}m\frac 1{\rho(\BV x)}\parderiv{}{\BV
  x}\Big[\theta(\BV x)\rho^2(\BV x)
  \int_0^\infty\left(1-e^{-\frac{\phi(r) }{\theta(\BV
      x)}}\right)\parderiv{}r\big(r^3Y(\theta(\BV x),r)\big)\dif
  r\Big],
\end{equation}
which is what appears in \eqref{eq:bphi}.

\subsection{Collision integral}

Changing the dummy variable of integration $\BV x_2=\BV x-\sigma\BV r$
in the collision integral of \eqref{eq:transport}, we arrive at
\begin{multline}
\frac 1m\int e^{-\frac{\phi(\|\BV x-\BV x_2\|)}{\theta((\BV x+\BV
    x_2)/2)}} Y_K(\theta((\BV x+\BV x_2)/2),\|\BV x- \BV
x_2\|)\lambda(\BV z-\BV z_2)\big(f(\BV z'')f(\BV z_2'')-f(\BV z)f(\BV
z_2)\big)\dif\BV z_2\\=\frac{\sigma^3}m\int\left[f(\BV x,\BV y,\BV
  v'',\BV w'')f\left(\BV x-\frac{\sigma\BV r}2,\BV y_2,\BV v_2'',\BV
  w_2''\right)-f(\BV x,\BV y,\BV v,\BV w)f\left(\BV x-\frac{\sigma\BV
    r}2,\BV y_2,\BV v_2,\BV w_2\right)\right]\\\lambda(\BV r,\BV y-\BV
y_2,\BV v-\BV v_2,\BV w-\BV w_2)e^{-\frac{\phi(\|\BV r\|)}{\theta(\BV
    x-\sigma\BV r/2)}} Y_K(\theta(\BV x-\sigma\BV r/2),\|\BV
r\|)\dif\BV r\dif\BV y_2\dif\BV v_2\dif\BV w_2,
\end{multline}
where the collision mappings are computed via \eqref{eq:gh_rescaled}.
Taking the limit as $\sigma\to 0$, $K\to\infty$, and $\sigma^3/m\sim$
constant, we further obtain
\begin{multline}
\frac{\sigma^3}m\int\lambda(\BV r,\BV y-\BV y_2,\BV v-\BV v_2,\BV w-
\BV w_2)e^{-\frac{\phi(\|\BV r\|)}{\theta(\BV x)}} Y(\theta(\BV x),
\|\BV r\|)\\\big(f(\BV x,\BV y,\BV v'',\BV w'')f(\BV x,\BV y_2,\BV
v_2'',\BV w_2'')-f(\BV x,\BV y,\BV v,\BV w)f(\BV x,\BV y_2,\BV v_2,\BV
w_2)\big)\dif\BV r \dif\BV y_2\dif\BV v_2\dif\BV w_2\\=
\frac{\sigma^3}m\int\alpha(\BV x,\BV r,\BV y-\BV y_2,\BV v-\BV v_2,\BV
w-\BV w_2)\\\big(f(\BV x,\BV y,\BV v'',\BV w'')f(\BV x,\BV y_2,\BV
v_2'',\BV w_2'')-f(\BV x,\BV y,\BV v,\BV w)f(\BV x,\BV y_2,\BV v_2,\BV
w_2)\big)\dif\BV r\dif\BV y_2\dif\BV v_2\dif\BV w_2,
\end{multline}
which is what appears in \eqref{eq:collision_integral}, with $\alpha$
given via \eqref{eq:alpha}.

\section{The entropy inequality for the Boltzmann--Vlasov equation}
\label{app:h-theorem}

The computation of \eqref{eq:lnf_inequality} proceeds as follows:
\begin{multline}
\langle\ln f\rangle_\coll(t,\BV x)=\frac
12\int\alpha\ln\left(\frac{f(\BV x,\BV y,\BV v',\BV w')f(\BV x,\BV
  y_2,\BV v_2',\BV w_2')}{f(\BV x,\BV y,\BV v, \BV w)f(\BV x,\BV
  y_2,\BV v_2,\BV w_2)}\right)\\f(\BV x,\BV y,\BV v, \BV w)f(\BV x,\BV
y_2,\BV v_2,\BV w_2)\dif\BV r\dif\BV y_2\dif\BV v_2 \dif\BV w_2\dif\BV
y\dif\BV v\dif\BV w\\=\frac 12\int\alpha\ln\left(\frac{f(\BV x,\BV
  y,\BV v',\BV w')f(\BV x,\BV y_2,\BV v_2',\BV w_2')}{f(\BV x,\BV
  y,\BV v, \BV w)f(\BV x,\BV y_2,\BV v_2,\BV w_2)}\right)\frac{f(\BV
  x,\BV y,\BV v, \BV w)f(\BV x,\BV y_2,\BV v_2,\BV w_2)}{f(\BV x,\BV
  y,\BV v',\BV w')f(\BV x,\BV y_2,\BV v_2',\BV w_2')}\\f(\BV x,\BV
y,\BV v',\BV w')f(\BV x,\BV y_2,\BV v_2',\BV w_2')\dif\BV r\dif\BV
y_2\dif\BV v_2 \dif\BV w_2\dif\BV y\dif\BV v\dif\BV w\\\leq\frac
12\int\alpha\left(1-\frac{f(\BV x,\BV y,\BV v,\BV w)f(\BV x,\BV
  y_2,\BV v_2,\BV w_2)}{f(\BV x,\BV y,\BV v',\BV w')f(\BV x,\BV
  y_2,\BV v_2',\BV w_2')}\right)\\f(\BV x,\BV y,\BV v',\BV w')f(\BV
x,\BV y_2,\BV v_2',\BV w_2')\dif\BV r\dif\BV y_2\dif\BV v_2 \dif\BV
w_2\dif\BV y\dif\BV v\dif\BV w=0.
\end{multline}

\end{document}